\documentclass[fleqn,10pt]{wlscirep}
\usepackage[utf8]{inputenc}
\usepackage[T1]{fontenc}
\usepackage{graphicx}% Include figure files
\usepackage{dcolumn}% Align table columns on decimal point
\usepackage{bm}% bold math
\usepackage{latexsym,epsfig}
\usepackage{graphicx}
\usepackage{verbatim}
\usepackage{comment}
\usepackage{amsmath}
\usepackage{amssymb}
\usepackage{stmaryrd}
\usepackage{color}
\usepackage{epstopdf}
\usepackage{grffile}
\usepackage{lipsum}
\usepackage{enumitem}
\usepackage{ulem}

\newcommand{\mrinal}[1]{\textcolor{blue}{\bf #1}}
\newcommand{\beq}{\begin{equation}}
\newcommand{\eeq}{\end{equation}}
\newcommand{\bea}{\begin{eqnarray}}
\newcommand{\eea}{\end{eqnarray}}
\newcommand{\ben}{\begin{eqnarray*}}
\newcommand{\een}{\end{eqnarray*}}
\newcommand{\bfig}{\begin{figure}}
\newcommand{\efig}{\end{figure}}

\renewcommand{\tt}{{\tilde{t}}}
\newcommand{\upa}{\uparrow}

\newcommand{\ua}{\uparrow}
\newcommand{\da}{\downarrow}
\title{Two component quantum walk in one-dimensional lattice with hopping imbalance}

\author[1]{Mrinal Kanti Giri}
\author[1]{ Suman Mondal}
\author[2,3,*]{Bhanu Pratap Das}
\author[1,2,*]{Tapan Mishra}
\affil[1]{Department of Physics, Indian Institute of Technology, Guwahati-781039, India}
\affil[2]{Centre for Quantum Engineering Research and Education,
TCG Centres for Research and Education in Science and Technology, Sector V, Salt Lake, Kolkata 70091, India}
\affil[3]{Department of Physics, School of Science, Tokyo Institute of Technology,
2-1-2-1-H86 Ookayama Meguro-ku, Tokyo 152-8550, Japan}
\affil[*]{mishratapan@iitg.ac.in, bpdas.iia@gmail.com}

\begin{abstract}
We investigate the two-component quantum walk in one-dimensional lattice. We show that the inter-component interaction strength together with the hopping imbalance between the components exhibit distinct features in the quantum walk for different initial states. When the walkers are initially on the same site, both the slow and fast particles perform independent particle quantum walks when the interaction between them is weak. However, stronger inter-particle interactions result in quantum walks by the repulsively bound pair formed between the two particles. For different initial states when the walkers are on different sites initially, the quantum walk performed by the slow particle is almost independent of that of the fast particle, which exhibits reflected and transmitted components across the particle with large hopping strength for weak interactions. Beyond a critical value of the interaction strength, the wave function of the fast particle ceases to penetrate through the slow particle signalling a spatial phase separation. However, when the two particles are initially at the two opposite edges of the lattice, then the interaction facilitates the complete reflection of both of them from each other. 
We analyze the above mentioned features by examining various physical quantities such as the on-site density evolution, two-particle correlation functions and transmission coefficients.
\end{abstract}
\begin{document}

\flushbottom
\maketitle
\thispagestyle{empty}

\section*{Introduction}

The dynamical evolution of isolated quantum many-body systems has been a topic of great interest in recent years in the context of non-equilibrium physics. The long time evolution of quantum states provides useful insights about the route to equilibration which is fundamentally very important to study physics related to localization and thermalization. While the time evolution of a many-body state provides actual dynamical behaviour of a system, quantum walk (QW) on the other hand, is an extremely versatile approach to address the dynamical behaviour of interacting systems at the few particle levels.

The phenomenon QW is the quantum analog of 
classical random walk which deals with the stochastic evolution of quantum walker(s) on a 
graph \cite{Aharonov}. Due to their relevance in fundamental physics by advancing our understanding of the quantum dynamics of different systems and possible applications in quantum technologies, QWs have attracted enormous attention in recent years~\cite{Kempe2003,Venegas-Andraca2012,ambainis2003quantum,Kempe_application,Childs_walk,childs2013universal,Schuld_2014,Ryan_QW}
The rapid experimental progress in the last decades have led to the observation of QWs in different systems such as trapped ions, neutral atoms,  
photons in photonic lattices and waveguides, biological systems 
etc~\cite{Schmitz2009,Zahringer2010,Karski2009,Weitenberg2011,Fukuhara2013,Manouchehri2014,Hoyer2010,Mohseni2008,PhysRevLett.119.130501,PhysRevX.7.031023,PhysRevLett.114.140502} at the single particle level. This has further facilitated to study the dynamical properties in systems with disorder, frustration and topological features~\cite{chalabi2019synthetic,schreiber2011decoherence,PhysRevA.86.063632,PhysRevA.101.032336,Eugene_et_al,PhysRevB.86.195414,wu2019topological, peng-xue2015,lozada-Vera2016}.

Considerable efforts have been made to investigate the role of interactions in the case of QWs for more than one indistinguishable particle in various physical contexts~\cite{Bromberg2009,Peruzzo2010,Sansoni2012,Broome2013,Spring2013,Tillmann2013,Crespi2013,Greiner_walk,Zakrzewski2017,Chaohong_etc,ahlbrecht2012molecular,mondalwalk,Siloi2017}. 
The combined effect of the inter-particle interaction and indistinguishability results in interesting 
features in different systems such as quantum gases in optical lattice ~\cite{Greiner_walk}, correlated photon pairs~\cite{Peruzzo2010,Lahini_walk,poulios2014quantum}, trapped ions~\cite{zahringer2010realization}, and superconducting qubits~\cite{Ye2019,Yan2019}. One such revelation is the spatial 
bunching of bosons in QWs due to the interaction between the two particles initially located at the same site and the Hanbury-Brown and Twiss (HBT) type interference when the two non-interacting bosons are located at two nearest neighbor sites~\cite{Greiner_walk,Lahini_walk}. In 
contrast, the presence of strong interactions between two nearest neighbour bosons leads to spatial 
anti-bunching due to fermionization~\cite{Greiner_walk,Lahini_walk}. These developments have paved the paths for quantum simulations  involving a few particles, and this provides a platform to have a bottom-up approach to understand the physics of many-body systems. Owing to their remarkable efficacy of probing many-body physics, QWs have been widely used to study different physical phenomena using both theoretical and experimental approaches~\cite{Greiner_walk,bloch_spin_charge_sep,Kitagawa2012_topological_QW, Ketterle2020, mondalwalk,Zakrzewski2017,sowinskiprb2020}. 

On the other hand, the physics of two component systems hosts a completely different scenario compared to the single particle systems. The combined role of inter and intra-component interaction, correlation and statistics play a crucial role in revealing novel physics which have been explored in great detail in the context of the Hubbard models. However, the experimental realization of such systems was made possible in systems of ultracold atoms in optical lattices. Considerable progress has been made 
in creating and manipulating binary atomic mixtures in optical lattices. Although the experiments using atomic mixtures are extremely complex compared to 
the single species systems, recent progress on the experimental front has made it 
possible to access Bose-Bose, Fermi-Fermi and Bose-Fermi mixtures in the
absence and presence of optical lattices~\cite{Taglieber,Fermi-Fermi,Ospelkaus2006,Tilman2006,Best,Catani,Gadway}. The complexities of such binary mixtures 
yield significant insights into the interacting spin model, atom-molecule interactions, 
quantum entanglement, topological phase transitions 
etc~\cite{Altman2003,Isacsson2005,Duan2003,Orth2008,Wang2016,Mathey2007,mishraps,Singh2017,SSHHexptLe2020,SantosSSH2018,Mondal2020,Ye2016}. Interestingly, the two component systems with hopping imbalance have  shown to reveal exciting new physics in various context~\cite{Altman2003, Duan2003, Mandel_et_al2003,Soltan-Panahi2011,Jian-wei-2017,Esslinger2015PRL}. Moreover, the dynamics of these constrained systems under proper conditions may reveal a different scenario in terms of transport properties and relaxation which has been investigated in a recent experiment~\cite{blochimbalance}. While the many-body simulations of such dynamical systems are challenging, the dynamics in the context of QW in such systems may reveal completely different physics which have gained attention in recent years~\cite{Roos2017,LindaSansoni2012,Aidelsburger2018,Sowinski_sarkar}.

Exploiting the experimental advances in the creation and manipulation of two-component quantum gases in optical lattices, in this work we study the physics characterizing the QW of a
two component system in a one-dimensional lattice. To this end we consider a system of two interacting particles of different hopping strength or different mass and show that the combined effect of hopping imbalance and interaction exhibits interesting physics in the two particle QW. Moreover, we show that the choice of initial states also plays an important role in the QW in such hopping imbalanced systems. Before going to the details of our studies, we briefly highlight the important findings which emerge from our analysis. We show that when the two particles start the QW from the same site, a repulsively bound pair~\cite{Winkler2006} is formed as a function of inter-particle interaction - a phenomenon similar to the case of two identical particles~\cite{Greiner_walk,Lahini_walk}. However, when the two particles start from two nearest neighbor sites, then the wavefunction of the fast component transmits through the slow component in the absence of interaction. With increase in interaction, the fast component completely gets reflected from the slower one before forming a weak doublon in the limit of weak interaction. However, when the walkers are few sites apart, the behaviour is similar to the previous case except that the doublon formation is not so prominent. Interestingly, when the two particles are initially located far from each other, then both the particles feel the effect of the interaction and reflect from each other.

\section*{Model and approach}
\label{sec:mm}
The Hamiltonian for the model which describes the system under consideration is given by;
%%%%%%%%%%%%%%%%%%%%%%%%%%%%%%%%%%%%%
\begin{eqnarray}
    H=-\sum_{\langle i,j\rangle,\sigma}J_\sigma(a_{i,\sigma}^{\dagger} a_{j,\sigma}+H.c.)+ U\sum_{i}n_{i,\da}n_{i,\ua}
%&+&\frac{W}{6}\sum_i n_i(n_i-1)(n_i-2)
\label{eq:ham}
\end{eqnarray} 
%%%%%%%%%%%%%%%%%%%%%%%%%%%%%%%%%%%%%
where $a_{i,\sigma}^{\dagger}$($a_{i,\sigma}$) is the creation(annihilation) operator of the two components denoted as $\sigma=\da,\ua$. 
$U$ is the inter-component interaction strength and $n_{i,\sigma}=a_{i,\sigma}^{\dagger} a_{i,\sigma}$ is 
the number operator at $i$'th site corresponding to each component $\sigma$. 
Here, $J_\sigma$ represents the nearest neighbor hopping matrix
element for the component $\sigma$. The two components are distinguished from each other by introducing the hopping imbalance in the system. For convenience we define $\delta=J_\da/J_\upa$ and 
the hopping imbalance in the system is 
incorporated by setting $\delta \neq 1 $. In our calculations, we consider $J_\ua > J_\da$ and set $J_\ua=1$ as the energy scale which makes all the physical quantities dimensionless. Due to the presence of one particle from each component,  the quantum statistics of individual components can be neglected. 

Our studies are based on the continuous-time quantum walk (CTQW) approach~\cite{Farhi1998,Kempe_2003,Childs_2002} which is based on the dynamical evolution of an initial state under the influence of a time independent Hamiltonian as shown in Eq.\ref{eq:ham} as
\begin{equation}
    |\Psi(t)\rangle=e^{-iHt/\hbar}|\Psi_{0}\rangle
\end{equation}
where, $|\Psi_0\rangle$ is the initial state. For our studies we consider different initial states depending upon the initial positions of the particles. Hereafter, we refer to the CTQW as only QW for convenience. 

In order to understand the physics of the system, we primarily compute two 
important physical quantities such as the expectation value of the on-site number operator as
\begin{equation}
 \langle n_i(t)\rangle =\langle \Psi(t)| \sum_\sigma a_{i,\sigma}^{\dagger}a_{i,\sigma}|\Psi(t)\rangle
 \label{eq:ni}
\end{equation}
and the two particle correlation function  
\begin{equation}
 \Gamma_{ij} = \langle a_{i,\ua}^\dagger a_{j,\da}^\dagger a_{j,\da} a_{i,\ua}\rangle
 \label{eq:gamma}
\end{equation}
with the time evolved state $|\Psi(t)\rangle$. 
Note that $\Gamma_{ij}$ defined here as the correlation function between the two components and 
is different from the two-particle correlation function defined in Refs.~\cite{Lahini_walk,Greiner_walk,mondalwalk} for identical particles.
For our analysis we compute $\Gamma_{ij}$ after an evolution time, $t$. Apart from these two important observables, we also analyze other quantities of interest such as the half-length occupation, point of contact and transmission coefficients which we describe in the following section.
In our numerical simulations, we consider a lattice of length $L=41$ with open boundary condition such that we have $20$ sites in the left and right of the central sites with index \textquotedblleft$0$\textquotedblright. 
In all the cases, we study the QWs by varying $U$ from zero to a large repulsive limit. Note that similar physics is expected for attractive interactions as well. By considering different values of $\delta$ for different initial states, we study the QWs as discussed in detail in the following section. 
%\tapan{Due to the symmetry of the model, the 
%results for the attractive $U$ is identical to the repulsive one. }

%%%%%%%%%%%%%%%%%%%%%%%%%%%%%%%%%%%%%%%%%%%%%%%%%%%%%%%%
\begin{figure}
	\begin{center}
		%\vspace*{-2mm}
		\includegraphics[width=0.5\columnwidth]{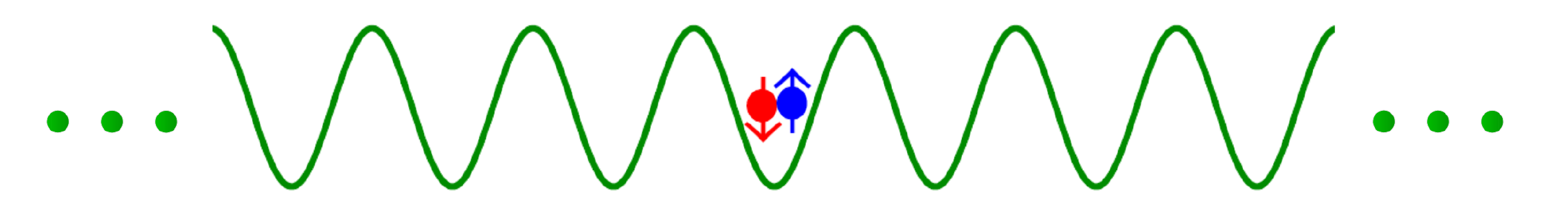}
	\end{center}
	%\vspace*{-4mm}
	\caption{The schematic description of the initial states given in Eq.~\ref{eq:psi0}.}
	\label{fig:latt_1}
\end{figure}
%%%%%%%%%%%%%%%%%%%%%%%%%%%%%%%%%%%%%%%%%%%%%%%%%%%%%%%%

\section*{Results}
\label{sec:re}

\subsection*{Two particles at the same site}
%$|\Psi_0\rangle=a_{0,\ua}^\dagger a_{0,\da}^\dagger|vac\rangle$} 
\label{ssec:1}
In this section, we start with the QW of $\ua$ and $\da$ particles which are initially located at the central site of the lattice as shown in Fig.~\ref{fig:latt_1}. The initial state corresponding to this situation is given as  ;
\begin{equation}\label{eq:psi0}
 |\Psi(0)\rangle=a_{0,\ua}^\dagger a_{0,\da}^\dagger|vac\rangle
\end{equation}
where, $|vac\rangle$ represents the empty state. Note that in 
the absence of any hopping imbalance i.e. $\delta=1$, 
the system is similar to that of two indistinguishable interacting particles whose QW has 
already been studied in detail~\cite{Lahini_walk,Greiner_walk}. It has been shown in both theoretical
and experimental analysis that 
when $\delta=1$, the two particles exhibit bosonic bunching as 
a function of interaction. 

%%%%%%%%%%%%%%%%%%%%%%%%%%%%%%%%%%%%%
% FIGURE 1
\begin{figure}[!h]
	%\begin{center}
	\includegraphics[width=0.5\columnwidth]{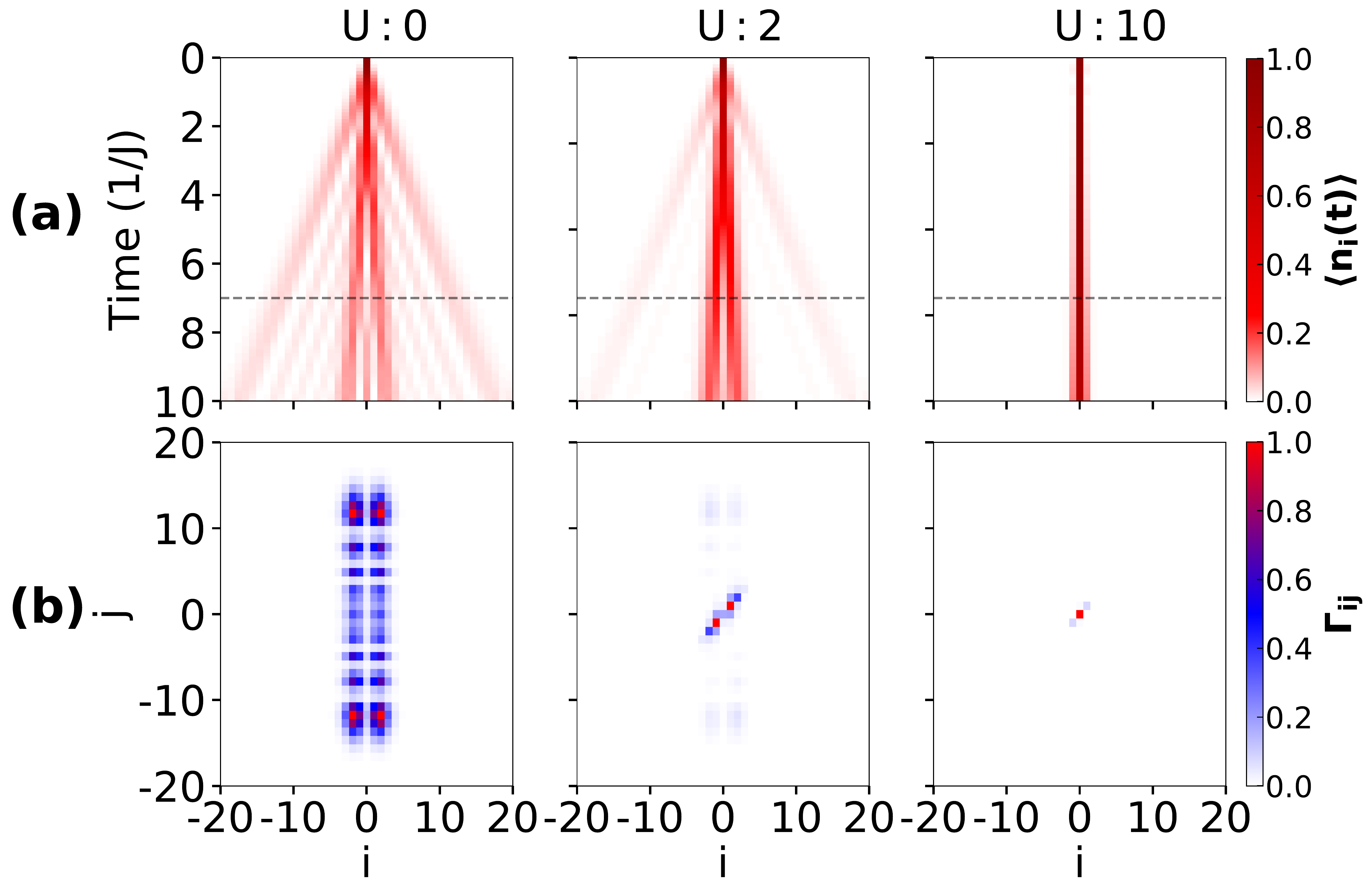}
	\includegraphics[width=0.5\columnwidth]{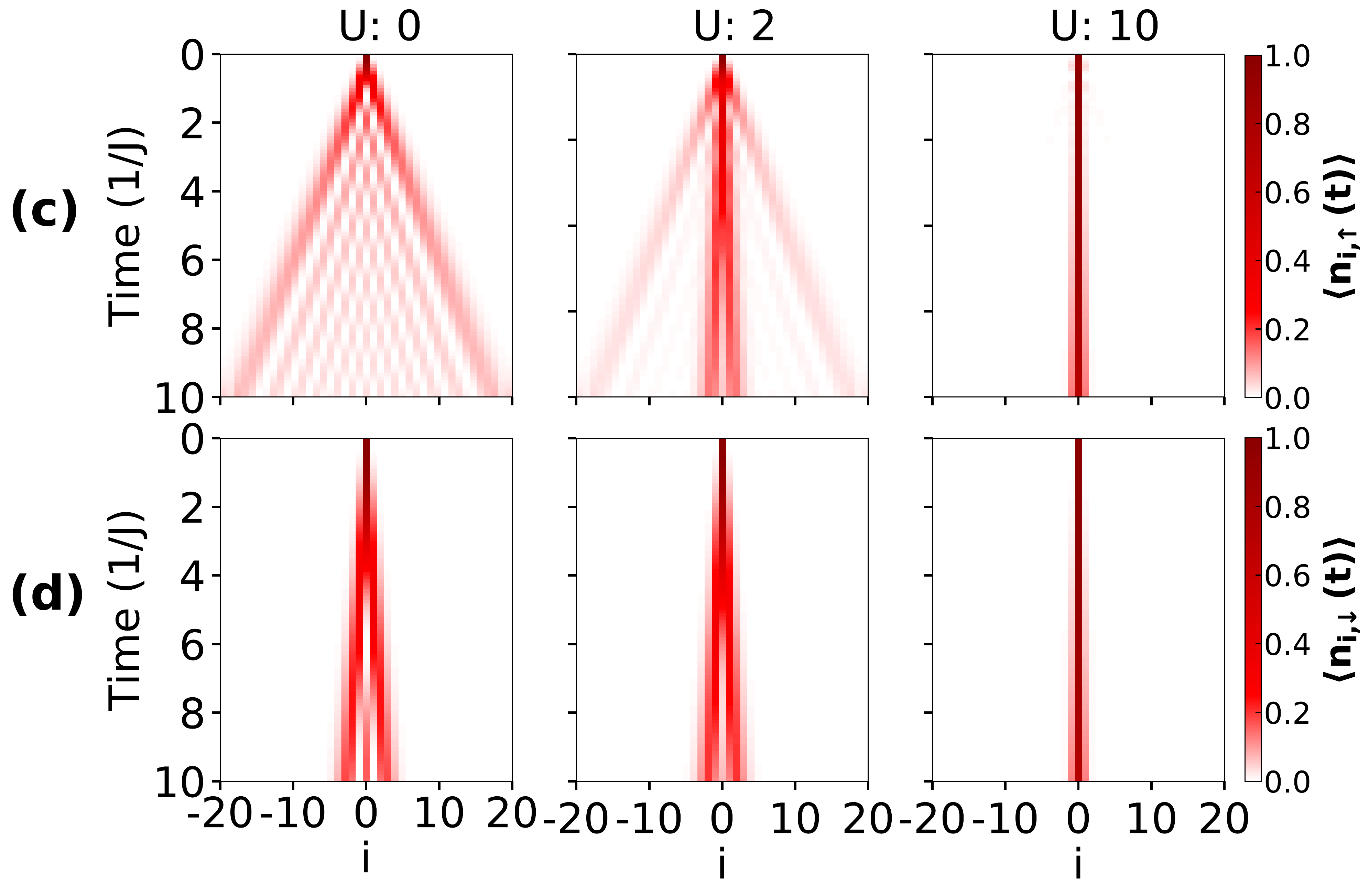}
	%\end{center}
	\caption{Figure shows the QW of two particles with the initial state given in Eq.~\ref{eq:psi0} and $\delta = 0.2$. 
		(a) Shows the time evolution of the normalized on-site density for different values of $U$. (b) Shows the normalized correlation functions $\Gamma_{ij}$ at time $t=7J^{-1}$, which correspond to the dashed lines in (a).
		(c) and (d) shows the on-site density evolution of $\ua$ particle and $\da$ particle respectively corresponding to the QW shown in (a).}
	\label{fig:ni1}
\end{figure}
%%%%%%%%%%%%%%%%%%%%%%%%%%%%%%%%%%%%%

However, in the present case the introduction of hopping imbalance i.e. $\delta \neq 1$, makes the particles distinguishable, which may exhibit different features in the QW. 
In this context, we first consider $\delta=0.2$ and vary the interaction strength $U$ and analyze the spreading of the on-site particle density, which is depicted in Fig.\ref{fig:ni1}(a). It can be seen that for vanishingly small interactions, the two particles exhibit independent particle QW. Due to the difference in hopping strength between the particles, the density profile of the $\da$ particle spreads at a slower rate compared to the $\ua$ particle, as expected. However, as the strength of interaction increases ($U=2$), the density profile exhibits simultaneous features of single and composite particle QW, a result similar to the ones discussed in Ref.~\cite{Greiner_walk, Zakrzewski2017}. Further increase in interaction to a large value results in only a single profile corresponding to a slow spreading of the density, indicating that the two particles performs QW as a composite object. This feature in the QW can be attributed to the formation of doublons ($\ua\da$) due to the large onsite interaction~\cite{Winkler2006} (see Fig.~\ref{fig:ni5}(a)). Hence, for stronger interaction, the QW of an effective doublon appears, which can be seen as a very slow evolution of the density profile in Fig.~\ref{fig:ni1}(a) for $U=10$.

This feature of doublon formation can be clearly seen by separately looking at the evolution of individual particle's on-site 
densities $\langle n_\sigma\rangle$ over the lattice.
% and also the densities at a particular instant of time as shown in  
% \suman{as well as the distribution at time $t=7J^{-1}$ as shown in Fig.~\ref{fig:ni5}(a)} 
% for $\ua$ and $\da$ particles. 
Clearly, with increasing 
$U$, the spreading of both the $\ua$ and $\da$ particles become slower and identical to each other for large values of $U$ as depicted in Fig.~\ref{fig:ni1}(c) and Fig.~\ref{fig:ni1}(d) respectively. 
%These features can also be seen in 
%the densities of the individual components $\langle n_\sigma\rangle$ at a particular instant ($t=7J^{-1}$) as shown
%Fig.~\ref{fig:ni5}(a). 
An accurate insight about this doublon formation 
can be understood by analyzing the two particle correlation matrix $\Gamma_{ij}$ defined in Eq.~\ref{eq:gamma}. 
We calculate $\Gamma_{ij}$ after evolving the initial state to $t=7J^{-1}$ (indicated by the
dashed line in Fig.~\ref{fig:ni1}(a)) and plot it in Fig.~\ref{fig:ni1}(b) for different $U$ considered in Fig.~\ref{fig:ni1}(a). 
At $U=0$, the two-particle correlation matrix shows four peaks at four different locations. 
This feature is different from the $\delta =1$ case where the four peaks 
appear at four symmetric positions~\cite{Lahini_walk} as the wave functions of each non-interacting particle spreads the same distance from the center on either side. However, due to the hopping imbalance, the spreading of the wave functions is not identical for the two particles, and this results in an asymmetry in the position of the peaks in the two-particle correlation matrix. By increasing $U$, the diagonal part of the matrix start to dominate, and eventually, for large $U$, only the dominating diagonal part survives, which indicates the formation of doublon (see Fig.~\ref{fig:ni1}(b)).
 %%%%%%%%%%%%%%%%%%%%%%%%%%%%%%%%%%%%%
% FIGURE 
\begin{figure}[!b]
%\begin{center}
	%\vspace*{-2mm}
	%\includegraphics[width=.95\columnwidth]{ni1corr.pdf}
	\includegraphics[width=0.45\columnwidth]{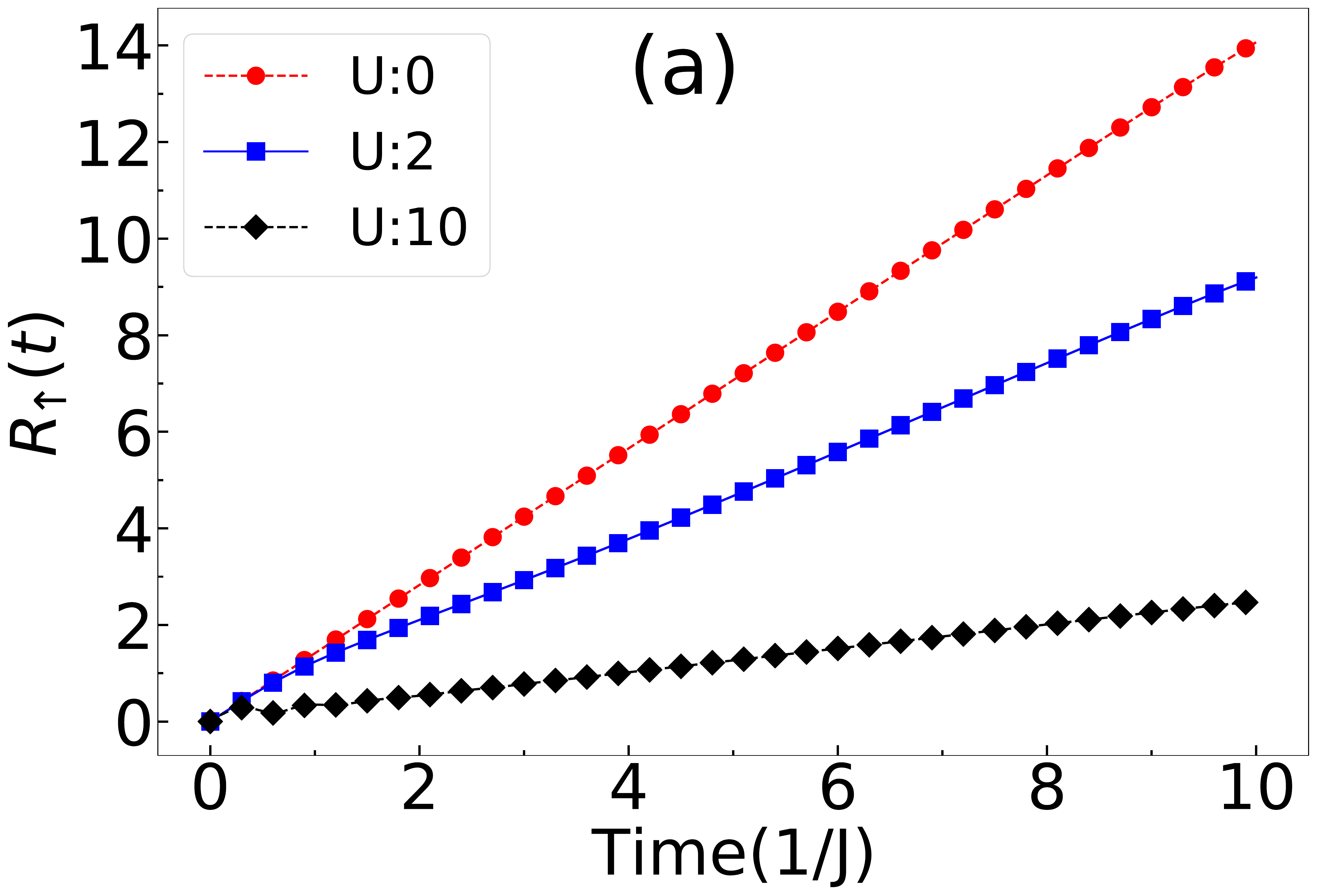}
	\hspace{6mm}
	\includegraphics[width=0.45\columnwidth]{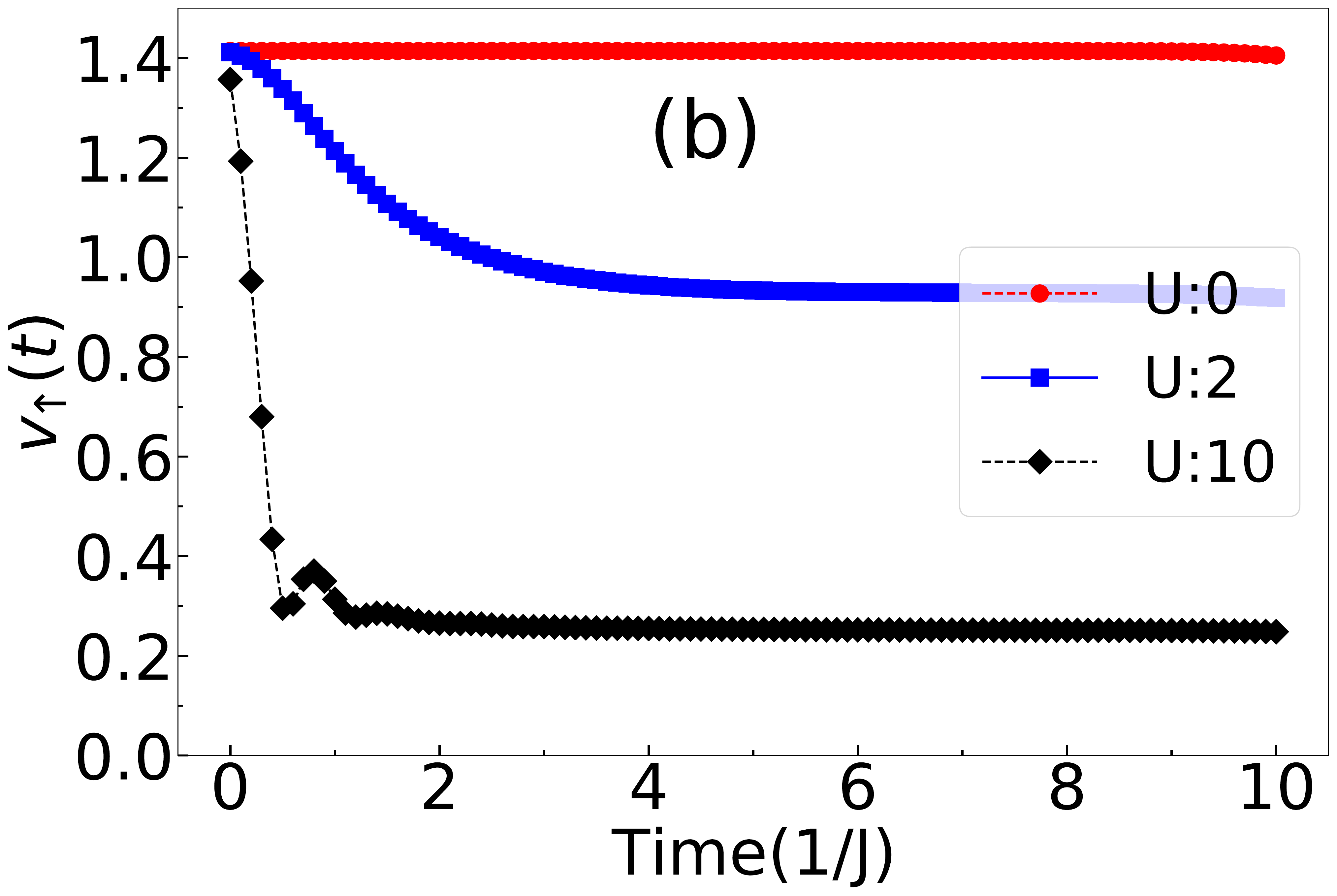}
%\end{center}
\vspace*{-2mm}
\caption{(a) $R(t)$ and (b) $v(t)$ are plotted for $U=0,~2$ and $10$ corresponding to the expansion of the wave function of $\ua$ particle when $\delta=0.2$. }
\label{fig:vel}
\end{figure}
%%%%%%%%%%%%%%%%%%%%%%%%%%%%%%%%%%%%% 

\begin{figure}[!t]
	\begin{center}
		%\vspace*{-1mm}
		\includegraphics[width=0.5\columnwidth]{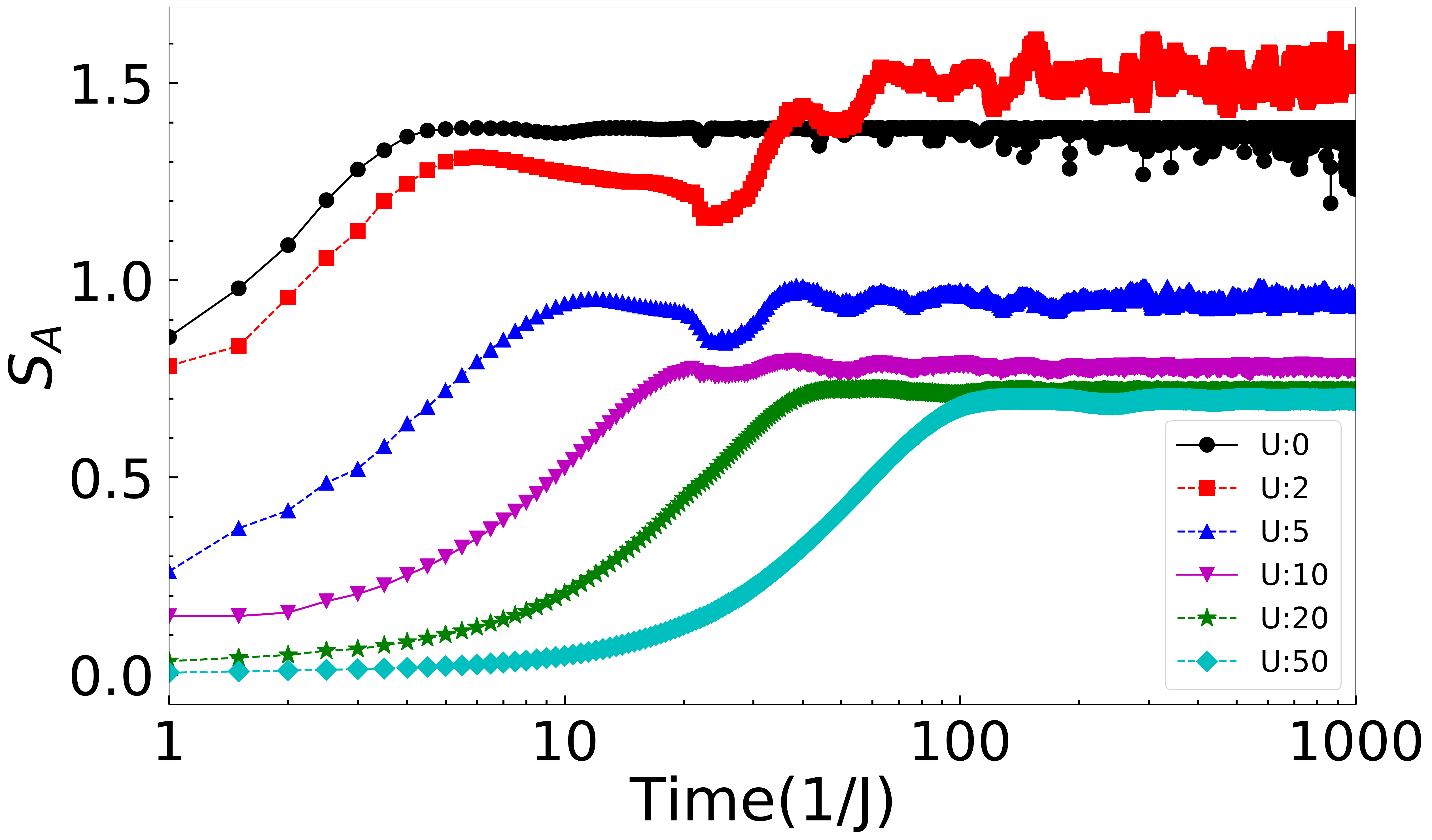}
	\end{center}
	%\vspace*{-4mm}
	\caption{Figure shows the time evolution of $S_A(t)$ for $U=0,~2,~5,~10,~20,~50$. }
	\label{fig:ent}
\end{figure}

To further complement the doublon formation we track the wave function expansion velocity  as 
\begin{equation}
v_\sigma(t)=R_\sigma(t)/t ,~~  \rm{where}~~ R_\sigma(t)=\left[\sum_i(i-i_0)^2\langle n_{i,\sigma}(t)\rangle\right]^{1/2}, 
\end{equation}
is the root mean-square displacement of the wave function and $i_0$ is the central site. In the limit of strong imbalance, the expansion of $\da$ particle slow. Hence, to check the slowing down of the composite system we plot $R_\ua(t)$ and $v_\ua(t)$ respectively of the $\ua$ particle wave function for different values of $U=0,~2,~10$ in Fig.~\ref{fig:vel}(a) and Fig.~\ref{fig:vel}(b) respectively. It can be seen for $U=0$, the expansion is fast which gradually slows down as $U$ increases. For $U=10$, the time evolution of $v_\ua(t)$ is extremely slow indicating the QW of bound pair with reduced effective hopping proportional to $J_\ua J_\da/U$.
Although the slow spreading of the wavefunction indicates a possible localization transition ~\cite{scirep_3B,Zakrzewski2017}, we rule out this possibility by computing the entanglement entropy defined as 
\begin{equation}\label{ent}
 S_A(t)=-Tr[\rho_A(t)~ln\rho_A(t)]
\end{equation}
by dividing the system into two equal subsystems $A$ and $B$ and computing the reduced density matrix $\rho_A(t)$ as 
\begin{equation}
 \rho_A(t)=Tr_B(|\psi(t)\rangle\langle\psi(t)|).
\end{equation}
We plot the time evolution of $S_A(t)$ for different values of $U=0,~2,~5,~10,~20,~50$ in Fig.~\ref{fig:ent} for $\delta=0.2$. The entanglement entropy grows initially for all values of $U$ but eventually saturates in the long time evolution indicating no localization~\cite{Zakrzewski2017}. 

 %%%%%%%%%%%%%%%%%%%%%%%%%%%%%%%%%%%%%%%%%%%%%%%%%%%%%%%%%%%%%%%%%%
%%%%%%%%%%%%%%%%%%%%%%%%%%%%%%%%%%%%%%%%%%%%%%%%%%%%%%%%%%%%%%%%%% 

Note that the feature of doublon formation is not due to the hopping imbalance, rather it is solely due to the inter-component interaction. 
However, the condition $\delta\neq1$ can influence the doublon formation due to the difference in kinetic energies between the particles.  
To further understand the effect of hopping imbalance, we check the QW for other values of 
$\delta$ such as $\delta=0.4,~0.6$ and $0.8$. For all the cases, the features 
in the QW remain qualitatively similar (not shown) but the signatures of doublon 
formation appear at stronger interaction strengths for larger values of $\delta$. 
To quantify the doublon formation we compute the quantity defined as 
\begin{equation}\label{eq:P}
 P = \sum_{i}\Gamma_{ii}=\sum_{i}n_{i,\da}n_{i,\ua}
\end{equation}
from the diagonal part of the two particle correlation matrix $\Gamma_{ij}$ during the time evolution.

\begin{figure}[!h]
%\begin{center}
	%\vspace*{-2mm}
	%\includegraphics[width=.95\columnwidth]{ni1corr.pdf}
	\includegraphics[width=0.45\columnwidth]{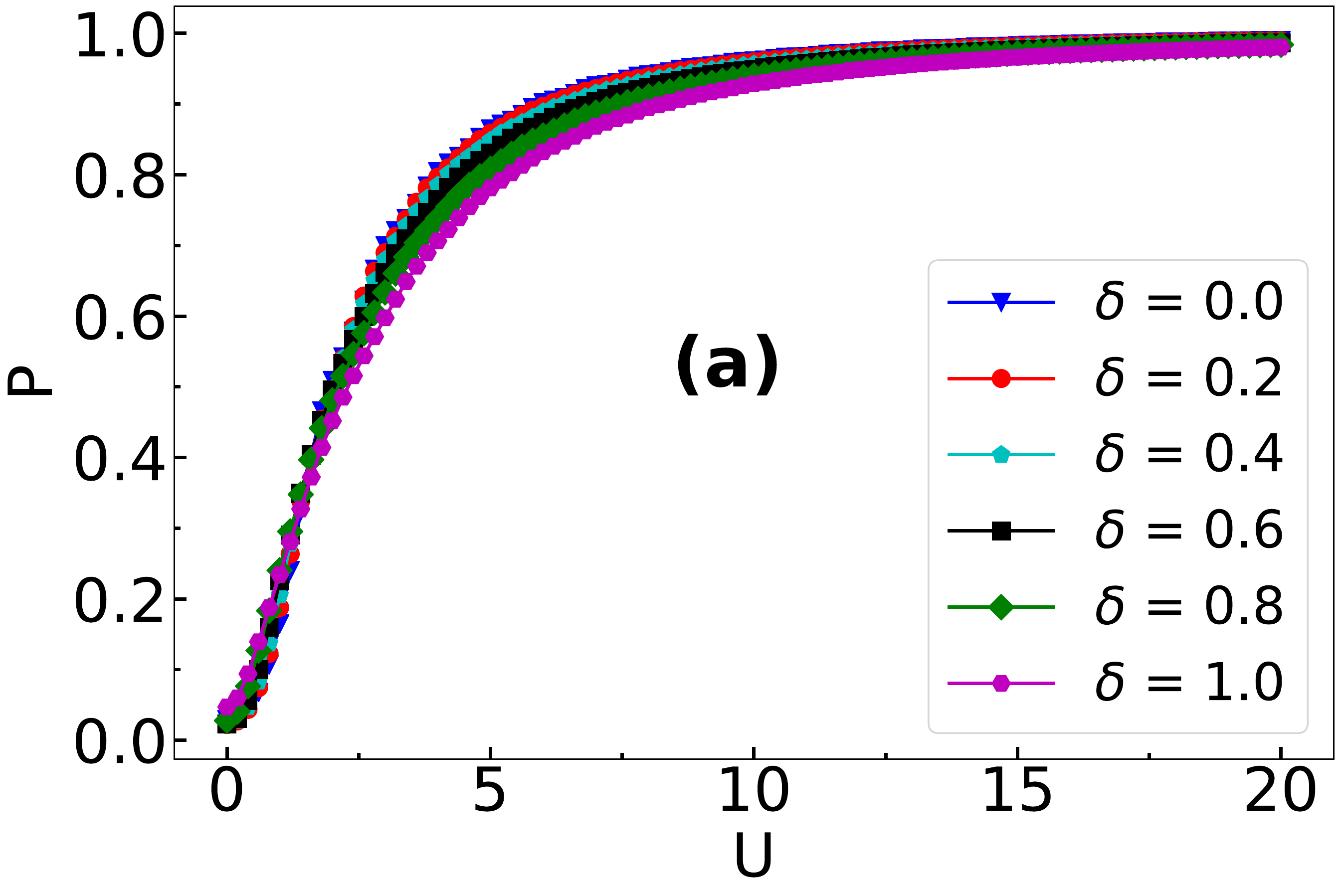}
	\hspace{6mm}
	\includegraphics[width=0.45\columnwidth]{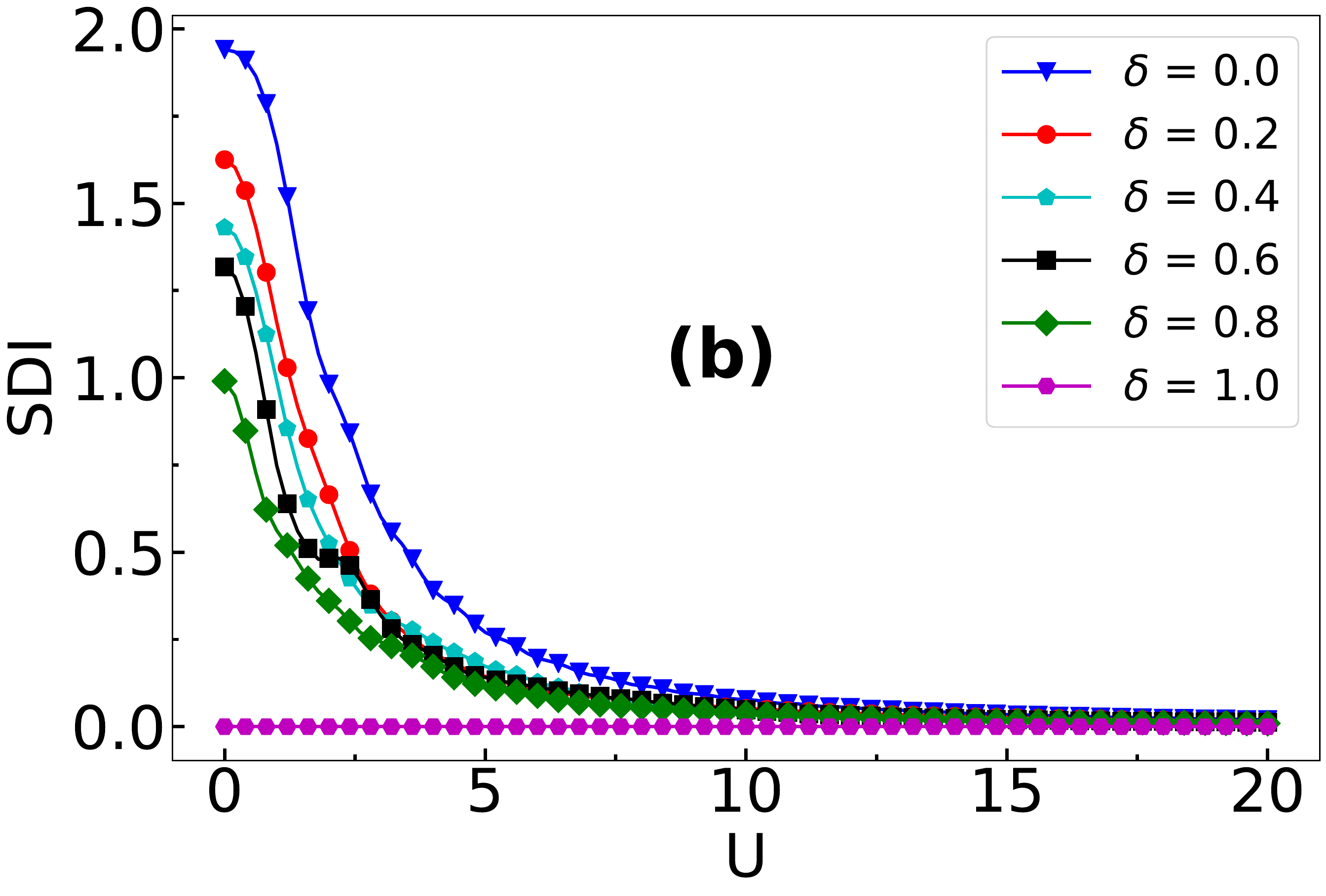}
%\end{center}
\vspace*{-2mm}
\caption{Figure shows the behaviour of (a) $P$ and (b) $SDI$ as a function of $U$ for different $\delta$ after a time evolution of the  initial state given in Eq.~\ref{eq:psi0} up to $t=7J^{-1}$. }
\label{fig:cor00}
\end{figure}

%after evolving the initial state up to certain time for different $U$. 
In our case, we compute $P$ at time $t=7J^{-1}$ for each values of $\delta$ and plot them as a function 
of $U$ in Fig.~\ref{fig:cor00}(a). The formation of doublons can be inferred from the behavior of $P$ which asymptotically 
approaches unity with increase in interaction strength. For comparison, we show $P$ for the two 
limiting cases i.e. $\delta=0$ and $1$ which correspond to the fully imbalanced and balanced cases respectively. 
From the figure it can be easily seen that although the effect of 
$\delta$ on the pair formation is not so significant, for strong imbalance (small $\delta$) the doublon formation happens at a smaller $U$ due 
to small effective hopping. 

%\tapan{Check species and replace them by components.}
To further complement the dependence of doublon formation on $\delta$ and $U$ we calculate the spatial density imbalance (SDI) between the two 
components which we define as  
\begin{equation}
 SDI = \sum_{i}|n_{i,\ua} - n_{i,\da}|.
\end{equation}
We plot the values of $SDI$ as a function of $U$ for different $\delta$ in Fig.~\ref{fig:cor00}(b), calculated at time $t= 7J^{-1}$ for the initial state given in Eq.~\ref{eq:psi0}. It can be seen that for all the cases of hopping imbalance, the values of $SDI$ are finite for smaller $U$ and eventually vanish in the regime of large $U$. While the vanishing of the $SDI$ for large $U$ is due to the doublon formation - a process similar to the balanced case ($\delta=1$), the finite values of $SDI$ for smaller values of $U$ can be attributed to the hopping imbalance. 
%Due to the hopping imbalance, the $\ua$ particle travels faster compared to the $\da$ one. Hence, both $\ua$ and $\da$ particles perform QW independently in the regime of small $U$. As a result, the density evolution of both the components are not identical (see Fig.~\ref{fig:ni5}(a)). 
% Due to the mass imbalance, the density evolution is not identical for both the component. Hence, S to the behaviour of $P$, in this case also the dependence of doublon formation  for all the values of $\delta$. In the small interacting regime, we see distinctions for different $\delta$ values. For the balanced case ($\delta = 1$), the $SDI=0$ since both the component travel the same distance. However, for the imbalanced case, with increasing the imbalance, the lighter particle wavefunction travels faster compared to the heavier one that results in a larger $SDI$. 
%\tapan{Check for QW and replace them by QW or vice-versa.}

%%%%%%%%%%%%%%%%%%%%%%%%%%%%%%%%%%%%%
% FIGURE 1
%%%%%%%%%%%%%%%%%%%%%%%%%%%%%%%%%%%%%

\subsection*{Two particles at two different sites}
\label{ssec:2}
\begin{figure}[h]
	\begin{center}
		%\vspace*{-1mm}
		\includegraphics[width=0.5\columnwidth]{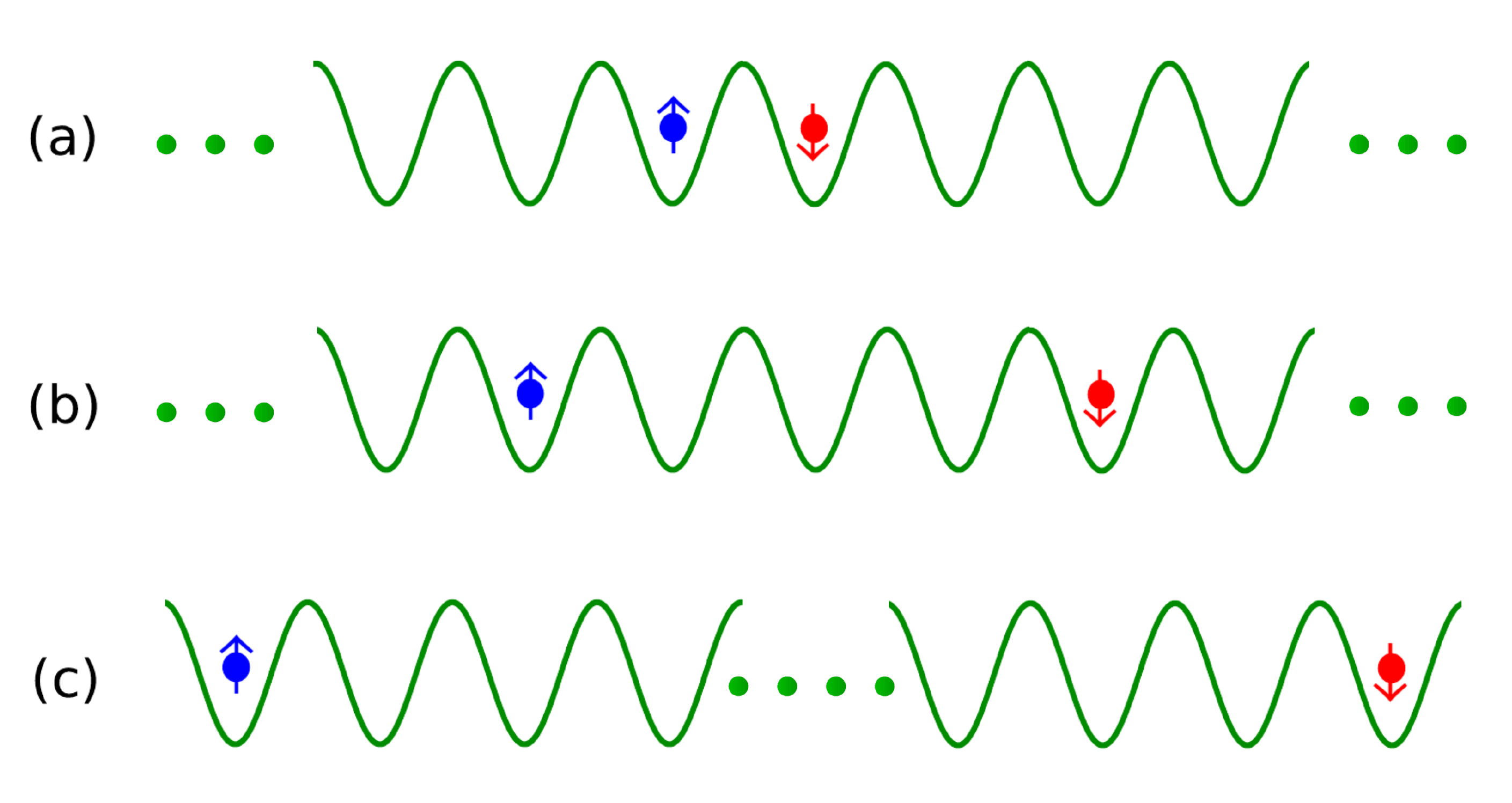}
	\end{center}
	%\vspace*{-4mm}
	\caption{(a), (b) and (c) depict the schematic description of the initial states given in Eq.~\ref{eq:psi1}, Eq.~\ref{eq:psi2} and Eq.~\ref{eq:psi3} respectively. }
	\label{fig:latt_all}
\end{figure}

In this section, we study the effect of hopping imbalance and interaction on the QW of two particles initially located at two different sites. We show that this situation reveals interesting physics as compared to the one discussed in the previous section where the effect of interaction was noticed in the form of doublon formation. 
To this end we consider different initial states which can describe various aspects of the QW at different parameter regime. 
In particular we consider three initial states which are given by
\begin{equation}\label{eq:psi1}
 |\Psi(0)\rangle=a_{0,\ua}^\dagger a_{1,\da}^\dagger|vac\rangle
\end{equation}
where the particles are at the nearest neighbor (Fig.~\ref{fig:latt_all}(a)),
\begin{equation}\label{eq:psi2}
 |\Psi(0)\rangle=a_{-2,\ua}^\dagger a_{2,\da}^\dagger|vac\rangle
\end{equation}
where there are three empty sites between the particles (Fig.~\ref{fig:latt_all}(b)) and 
\begin{equation}\label{eq:psi3}
 |\Psi(0)\rangle=a_{20,\ua}^\dagger a_{20,\da}^\dagger|vac\rangle
\end{equation}
where the particles are initially located at two edges of the lattice (Fig.~\ref{fig:latt_all}(c)). Although we have considered other initial states by varying the distance between the particles in our analysis, the above three states can reveal all the relevant physics. In the following we will mainly focus on the QW for all the three different cases mentioned above by analyzing various relevant physical quantities such as the evolution of density, correlation matrix and transmission coefficients. The results arising due to the other initial states will be highlighted when necessary. 
%%%%%%%%%%%%%%%%%%%%%%%%%%%%%%%%%%%%%%%%%%%%%%%%%%%%%%%%%%%%%%%%%%
%%%%%%%%%%%%%%%%%%%%%%%%%%%%%%%%%%%%%%%%%%%%%%%%%%%%%%%%%%%%%%%%%%
%\tapan{In all the figures we should mention $\langle n_i(t)\rangle$ instead of just $\langle n_i\rangle$ I think. }
\begin{figure}[t]
	\begin{center}
		%\vspace*{-20mm}
		\includegraphics[width=0.5\columnwidth]{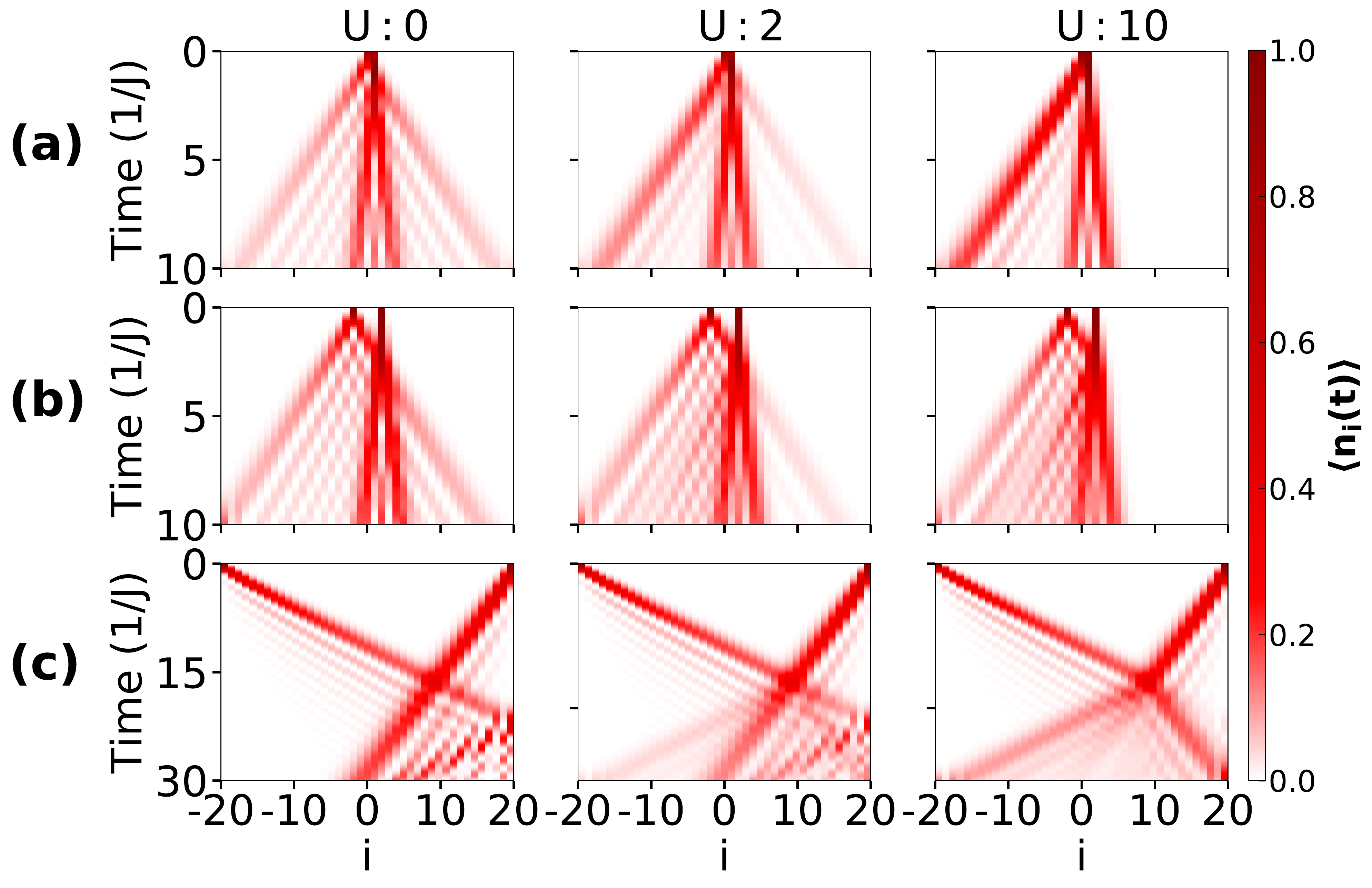}
	\end{center}
	%\vspace*{-2mm}
	\caption{Figure shows the QWs for different initial states and different values of $U$. Here (a), (b) and (c) depict the total density (normalized) evolution for the three initial states Eq.~\ref{eq:psi1}, Eq.~\ref{eq:psi2} and Eq.~\ref{eq:psi3} respectively. For results depicted in (a) and (b) $\delta=0.2$ and for (c) $\delta=0.4$ has been considered.}
	\label{fig:ni3}
\end{figure}
\subsubsection*{Density evolution}
First, we study the behaviour of the on-site densities in the two particles QW by considering different values of $\delta$ and by varying $U$. The time evolution of $\langle n_{i}\rangle$ with the initial states given in Eq.~(\ref{eq:psi1} - \ref{eq:psi3}) are depicted in Fig.~\ref{fig:ni3}(a-c) respectively. From the figure, one can see a marked difference compared to the situation where the two particles are initially located at the same site (see Fig.~\ref{fig:ni1}(a)). It can be noticed that 
there also exist some similarities between the two scenarios at vanishingly small interaction when both the components perform independent particle QWs and the $\ua$ particle (left) spreads faster compared to the $\da$ particle (right). For finite $U$, both the particles start to interact with each other after a certain time and position, leading to interesting features in the QW. 

\begin{figure}[!h]
	\begin{center}
		%\vspace*{-20mm}
		\includegraphics[width=0.6\columnwidth]{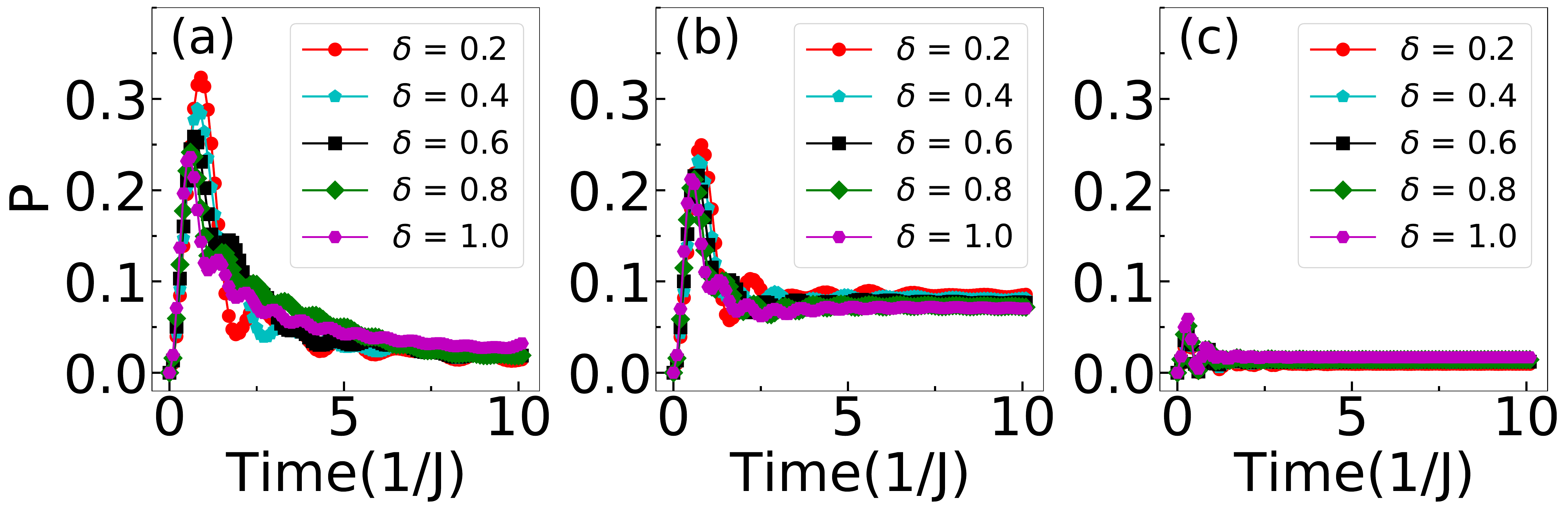}
	\end{center}
	%\vspace*{-2mm}
	\caption{Time evolution of $P$ for different values of $\delta$ and (a) for $U=0$, (b) for $U=2$   and (c) for $U=10$.}
	\label{fig:allP}
\end{figure}

When the two particles are initially located at the adjacent sites (Eq.~\ref{eq:psi1} and Fig.~\ref{fig:latt_all}(a)), for $U=0$ the $\ua$ and $\da$ particles spread independently of each other as can be seen from Fig.~\ref{fig:ni3}(a). For finite but weak $U=2$, two different profiles corresponding to slow and fast spreading appear in the QW. This situation indicates the contribution from both single and doublon density evolution~\cite{Zakrzewski2017,Greiner_walk}. Further increase in the $U$, the two particles reflect from each other and the situation is similar to the anti bunching of identical bosons~\cite{Lahini_walk,Greiner_walk}.  To clearly understand this behaviour we plot the time evolution of $P$ defined in Eq.~\ref{eq:P} for different values of $U=0,~2,~10$ in Fig.~\ref{fig:allP}(a), Fig.~\ref{fig:allP}(b) and Fig.~\ref{fig:allP}(c) respectively. Clearly, the probability of pair formation for $U=0$ and $U=10$ vanishes with time which remains finite for $U=2$. The initial growth of $P$ in each case is due to the finite overlap of the two wave functions.

\begin{figure}[!h]
	%\begin{center}
		\includegraphics[width=0.5\columnwidth]{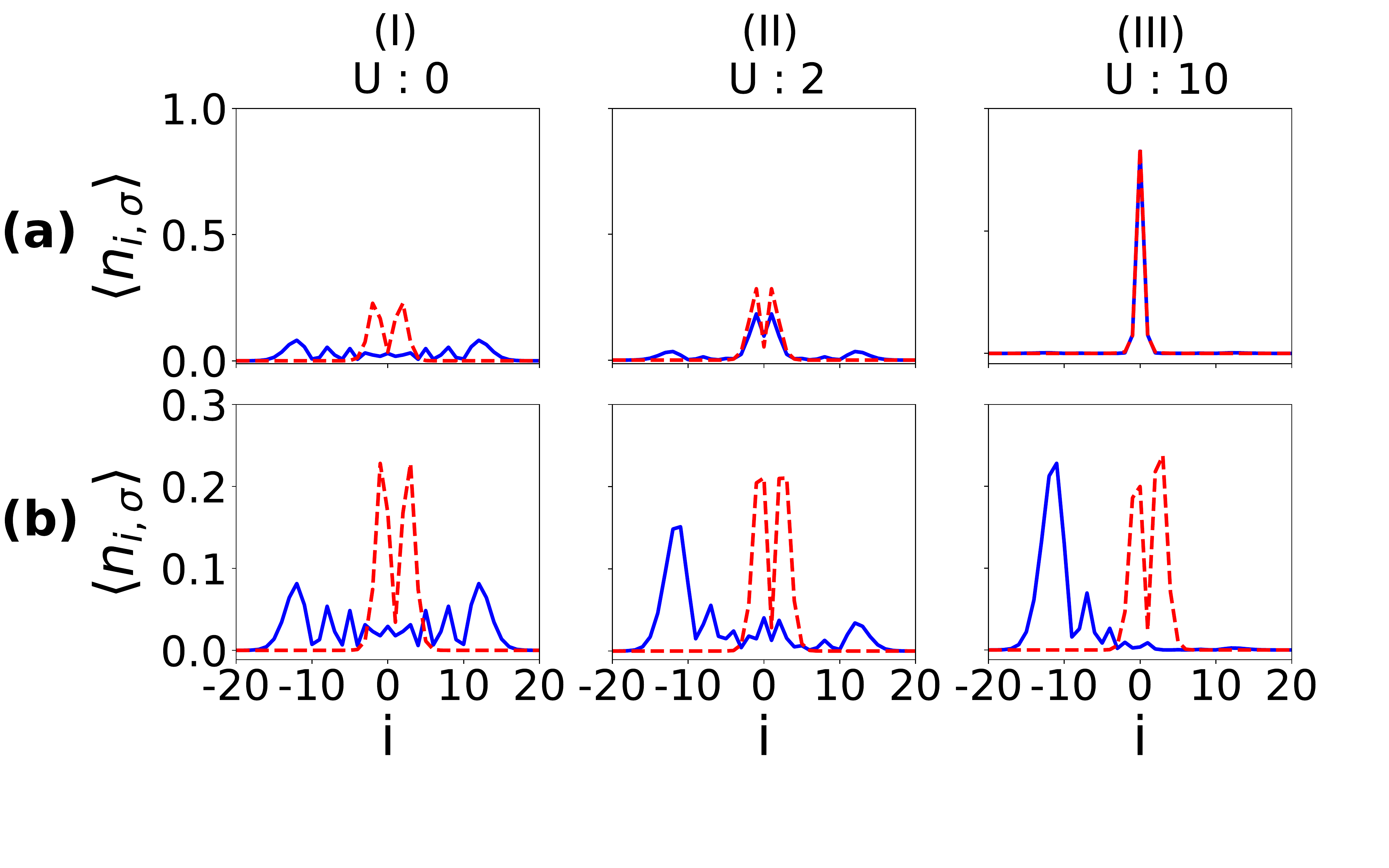}
	%	\vspace*{-20mm}
		\includegraphics[width=0.5\columnwidth]{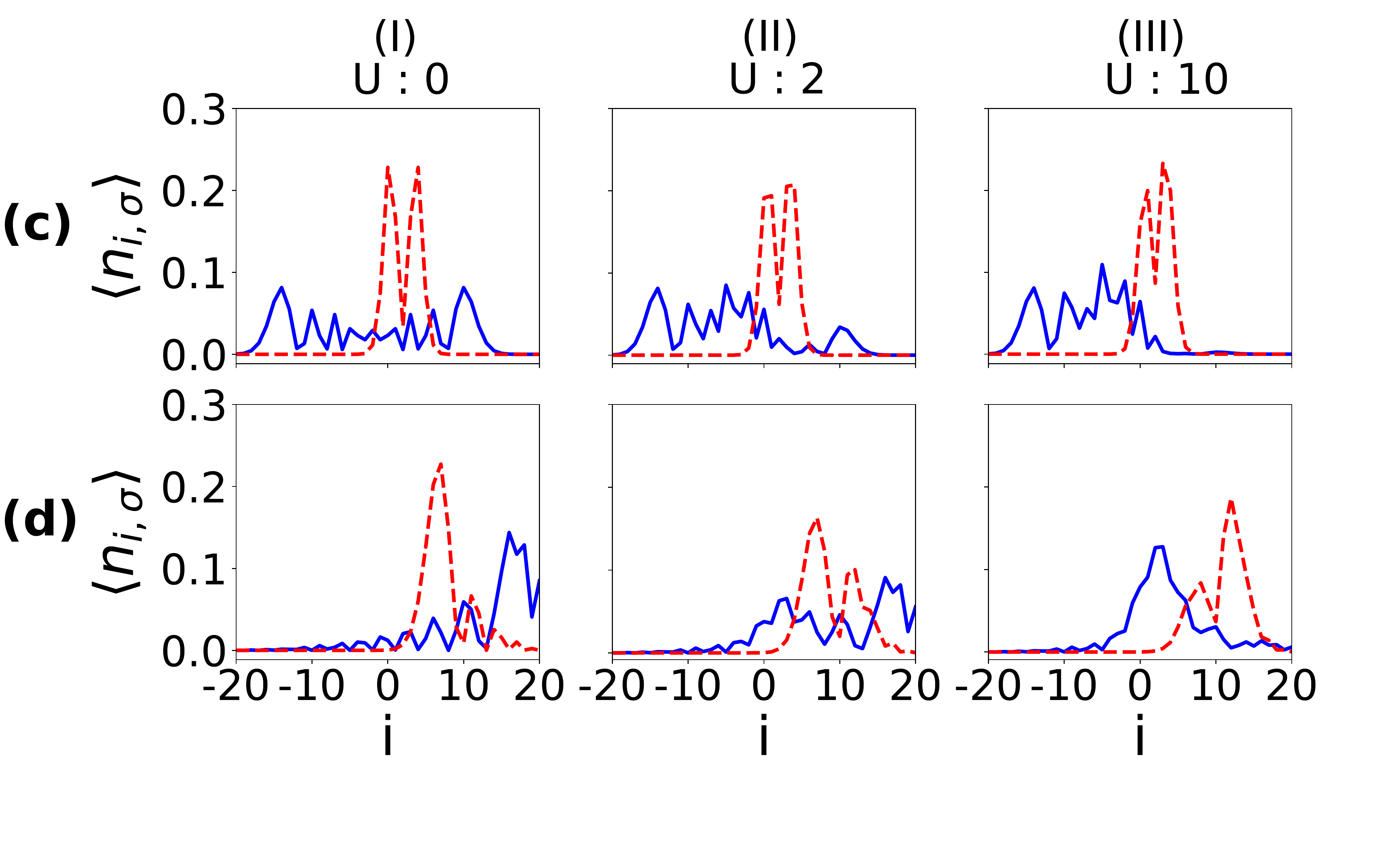}
	%\end{center}
%	\vspace*{-26mm}
	\caption{ Figure shows the onsite densities of $\da$ (dashed curves) 
		and $\ua$ particle (solid curves) in the lattice for different regimes of interaction 
		such as small (I), intermediate (II) and large (III) after evolving 
		the initial state up to a certain time ($t$). (a-d) correspond to 
		the initial states given in Eq.~\ref{eq:psi0}, Eq.~\ref{eq:psi1}, Eq.~\ref{eq:psi2} and Eq.~\ref{eq:psi3} respectively. For (a-c) $t=7J^{-1}$, $\delta=0.2$ and for (d) $t=20J^{-1}$, $\delta=0.4$ are considered.}%\tapan{Change site to ``i'' or keep ``site'' in all the figures.}}
	\label{fig:ni5}
\end{figure}

On the other hand, when the two particles are few sites apart ((Eq.~\ref{eq:psi2} and Fig.~\ref{fig:latt_all}(b)) and the interaction is finite but weak, the $\da$ particle acts like a barrier and as a result, the density spreading of the $\ua$ particle shows reflected as well as transmitted components in the propagation as shown in Fig.~\ref{fig:ni3}(b). As the interaction becomes stronger and stronger, the transmission ceases to occur and the $\ua$ particle wave function gets completely reflected for large enough $U$. Unlike the previous case, the pair formation is not stable during the time evolution (not shown). 
The QW for the initial state~(Eq.~\ref{eq:psi3} and Fig.~\ref{fig:latt_all}(c)), however, gives a very different outcome as can be seen from Fig.~\ref{fig:ni3}(c). Since the particles are initiated at the edges, we get a unidirectional spread of each particle's wave function. Due to the hopping imbalance, the density profiles of two particles meet at a point away from the center towards the slow moving particle ($\da$). When $U=0$, the two particles move independently and their wave functions transmit through each other without influencing the QWs of the individual particles. On the other hand, the onset of interaction $U$ leads to the reflection of both the components from each other by reducing the transmission. It can be easily seen that the effect of interaction on the $\da$ particle is drastic for this case compared to the other two cases. For clarity we also show the on-site density distribution over the entire lattice at a particular instant during the time evolution in Fig.~\ref{fig:ni5}(b-d) for the initial states shown in Eq.~(\ref{eq:psi1} - \ref{eq:psi3}). The effect of interaction can be clearly seen as we move from weak to strong interaction regime (I to III) in Fig.~\ref{fig:ni5}. For comparison we also show the situation when the two particles start from the central site in Fig.~\ref{fig:ni5}(a). 

The effect of interaction on the QW can be further understood by analyzing the evolution of the half-length occupation of the individual components which are defined as  %$N_{\frac{L}{2},\ua}$ and $N_{\frac{L}{2},\da}$%such as analyze the evolution of the number of particles occupying the left half of the system $N_{LA}$ and $N_{LB}$ with time, which are defined as 
%%%%%%%%%%%%%%%%%%%%%%%%%%%%%%%%%%%%%%%%%%%%%%%%%%%%%%%%%%%%%%%%%%%%%%%%%%%%%%%%
\begin{equation}
N_{\frac{L}{2},\ua} = \sum_{i\leq \frac{L}{2}} n_{i,\ua}
~ \text{and}~ 
N_{\frac{L}{2},\da} = \sum_{i\leq \frac{L}{2}} n_{i,\da}
\label{eq:Nla}
\end{equation}
for $\ua$ and $\da$ component respectively.

\begin{figure}[h!]
	%\begin{center}
	\hspace{2mm}
	\includegraphics[width=0.3\columnwidth]{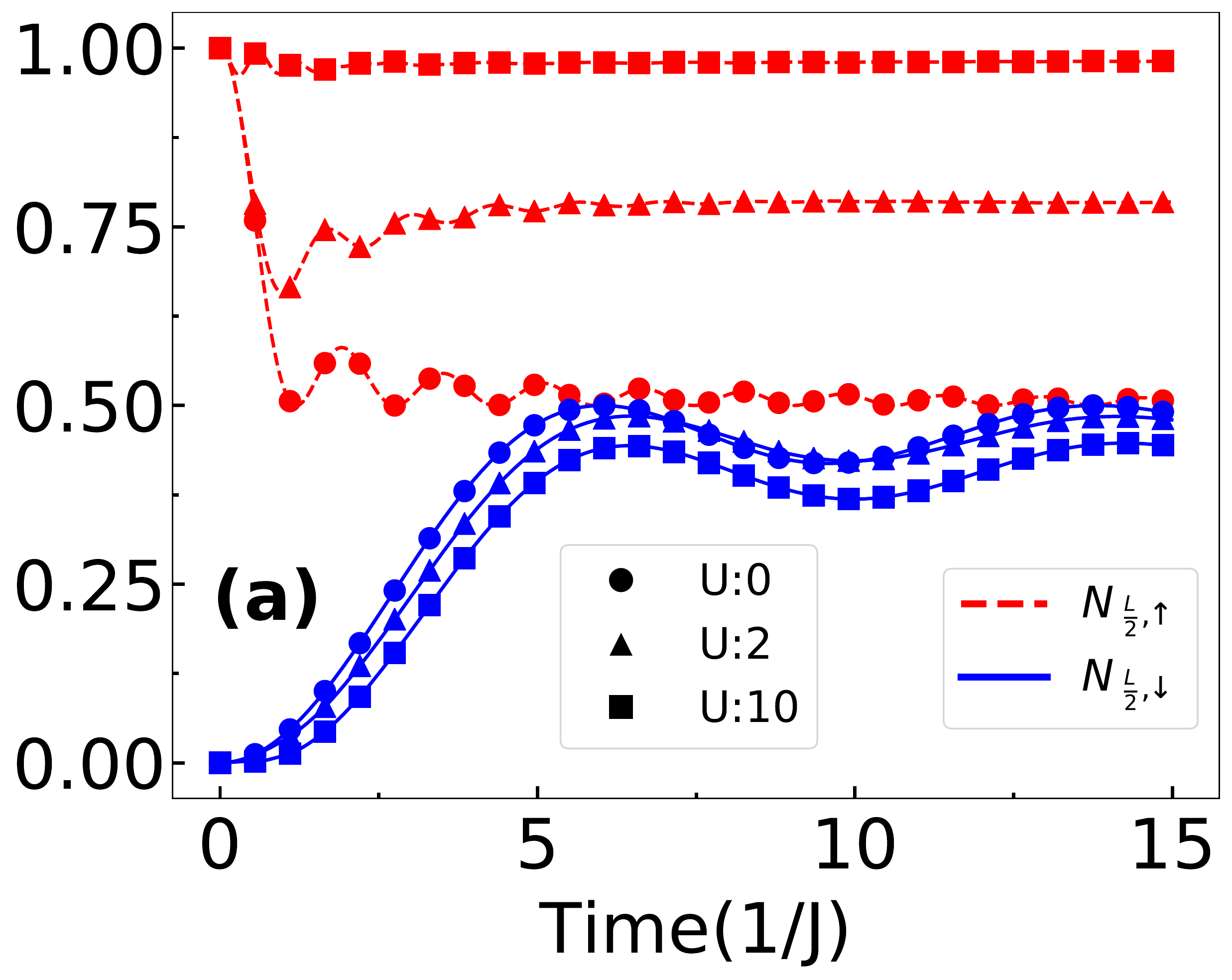}
	\hspace{3mm}
	\includegraphics[width=0.3\columnwidth]{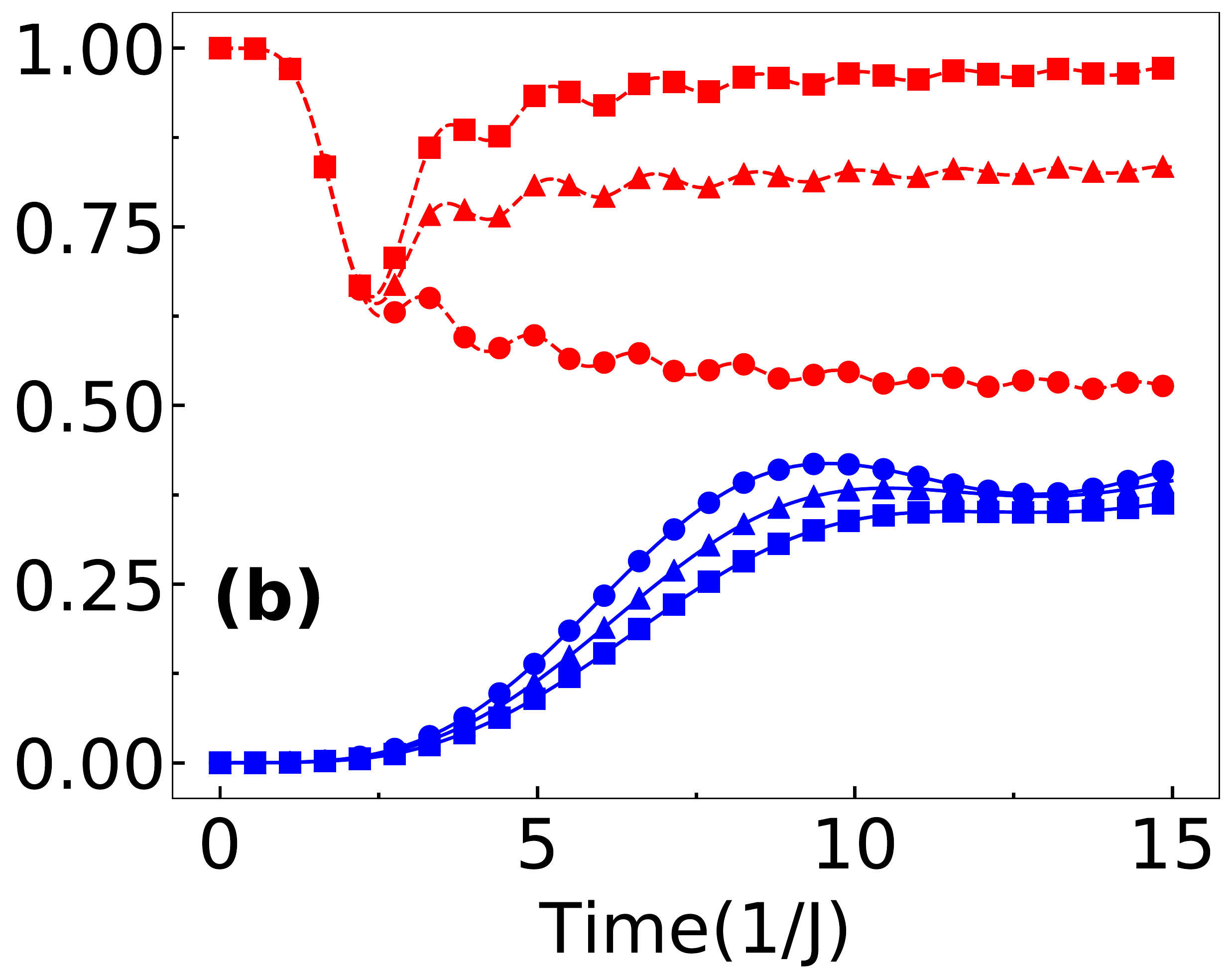}
	\hspace{3mm}
	\includegraphics[width=0.3\columnwidth]{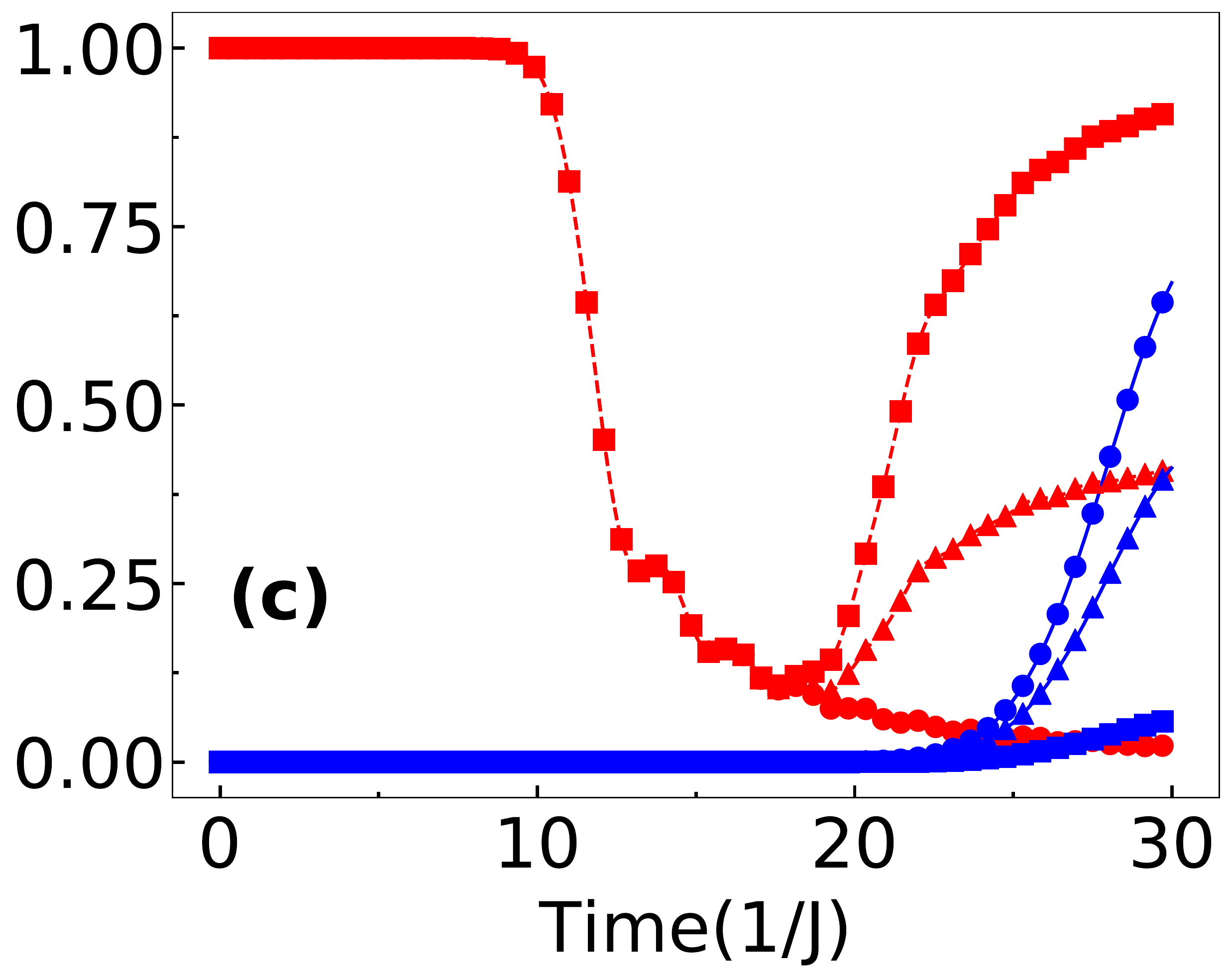}
	%\end{center}
	%\vspace*{-2mm}
	\caption{Evolution of half-length occupation $N_{\frac{L}{2},\ua}$ (dashed curves) and $N_{\frac{L}{2},\da}$ (solid curves) are shown for different interaction strengths such as $U=0$ (circles), $U=2$ (triangles) and $U=10$ (squares). (a), (b) and (c) correspond to the initial states given in Eq.~\ref{eq:psi1},~\ref{eq:psi2} and~\ref{eq:psi3} respectively. For (a-b) $\delta=0.2$ and for (c) $\delta=0.4$ is considered.}
	\label{fig:3nl}
\end{figure}
The time evolution of $N_{\frac{L}{2},\ua}$ (red dashed curves) and $N_{\frac{L}{2},\da}$ (blue solid curve) for different values of interactions such as $U=0$ (circles), $U=2$ (up triangles) and $U=10$ (squares) are plotted in Fig.~\ref{fig:3nl}(a-c) for the initial states and hopping imbalance considered in Fig.~\ref{fig:ni3}(a-c) respectively.
% \sout{shown in  Eq.~(\ref{eq:psi1}-\ref{eq:psi3}) respectively.} 
From the figures it can be seen that initially $N_{\frac{L}{2},\ua}=1$ and $N_{\frac{L}{2},\da}=0$ as the $\ua$ and the $\da$ particles reside in the left and the right halves of the system respectively. As the time progresses, different features are visible in the time evolution of $N_{\frac{L}{2},\ua}$ and $N_{\frac{L}{2},\da}$ for different initial states and interactions due to hopping imbalance. 

In Fig.~\ref{fig:3nl}(a), for $U=0$ the value of $N_{\frac{L}{2},\ua}$ ($N_{\frac{L}{2},\da}$) initially starts to decrease (increase) as both the wave functions transmit through each other. Eventually both the quantities saturate to a value close to $0.5$ due to no reflection from each other. Finite interactions however, lead to reflection of wave functions and hence $N_{\frac{L}{2},\ua}$ saturates to different values larger than $0.5$. For sufficiently strong $U$, $N_{\frac{L}{2},\ua}$ saturates to unity due to complete reflection from the $\da$ particle. These features can be seen from the curves corresponding to $U=2$ and $10$ in Fig.~\ref{fig:3nl}(a). Note that the effect on the $\da$ particle in this process is negligible. For the second case (Fig.~\ref{fig:3nl}(b)), while the long time evolution of $N_{\frac{L}{2},\ua}$ and $N_{\frac{L}{2},\da}$ exhibit features similar to the case shown in Fig.~\ref{fig:3nl}(a), the short time evolution behave differently. Up to $t\sim1J^{-1}$, the values of $N_{\frac{L}{2},\ua}$ ($N_{\frac{L}{2},\da}$) remain equal to $1(0)$. This is because of the presence of empty sites between the particles at $t=0$ for which the $\ua$ particle wave function remains entirely on the left half of the lattice before spreading into the right half after $t=1J^{-1}$. During this time, the occupation by the $\da$ particle on the left half of the lattice remains zero. After $t=1J^{-1}$, however, the values of $N_{\frac{L}{2},\ua}$ suddenly decrease up to  $t\sim 2J^{-1}$ and then start to increase for values of $U\neq 0$. The decrease in the values of $N_{\frac{L}{2},\ua}$ is due to the hopping imbalance for which the $\ua$ and $\da$ particle wave function interact at a point right from the center of the lattice. Hence, there is a finite propagation of the $\ua$ particle wave function towards the right half of the lattice leading to the decrease in $N_{\frac{L}{2},\ua}$. After $t=2J^{-1}$, the values of $N_{\frac{L}{2},\ua}$ saturate at higher values as already discussed. On the other hand the values of $N_{\frac{L}{2},\da}$ increase and saturate after $t=1J^{-1}$. For the case shown in Fig.~\ref{fig:3nl}(c), the features are similar to the one shown in Fig.~\ref{fig:3nl}(b) except that the saturation occurs at a later time due to the largest distance between the particles at the initial position. Note that in our analysis we don't analyze the physics for a very long time evolution. Hence, the contributions arising from reflections from the boundaries are ignored in all the cases except the last case where the quantum walkers are initially located at the edges.

\subsubsection*{Correlation function}
\begin{figure}
	\begin{center}
		\includegraphics[width=0.5\columnwidth]{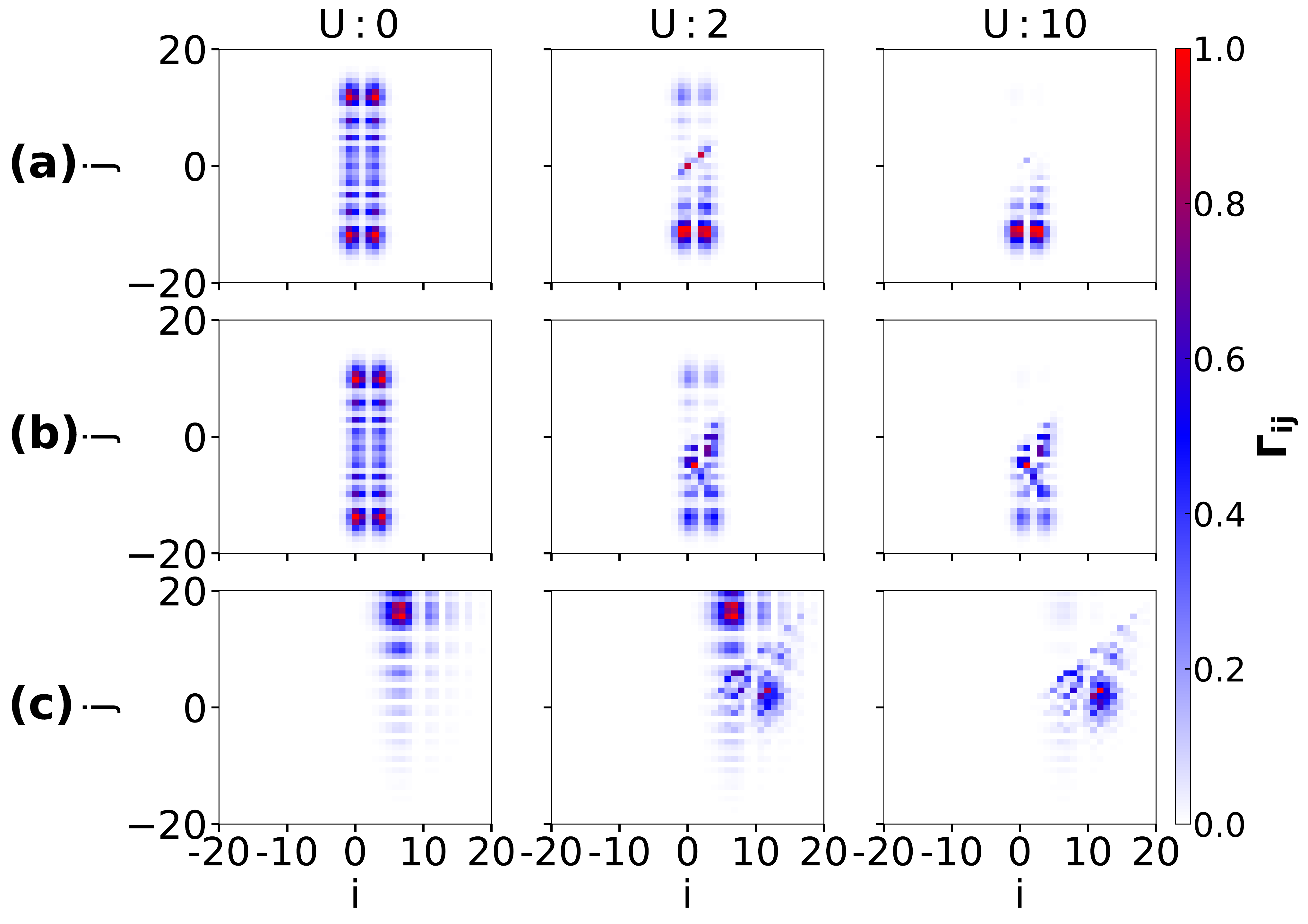}
	\end{center}
	\vspace*{-2mm}
	\caption{Normalized correlation functions  $\Gamma_{ij}$ are plotted corresponding to the parameters considered in Fig.~\ref{fig:ni3} at a particular instant during the time evolution. $\Gamma_{ij}$ in (a), (b) and (c) correspond to the initial states of Eq.~\ref{eq:psi1}, Eq.~\ref{eq:psi2} and Eq.~\ref{eq:psi3} respectively. While $\Gamma_{ij}$ is computed at $t=7J^{-1}$ for (a) and (b),  for (c) it is computed at $t=20J^{-1}$.}
	%\tapan{Either keep sites or i}.}
	\label{fig:cor2}
\end{figure}

The two-particle correlation function also shows interesting behavior due to the hopping imbalance and interaction. The $\Gamma_{ij}$ are computed for different values of $U$ considered in Fig.~\ref{fig:ni3} and plotted in Fig.~\ref{fig:cor2}(a-c) for the initial states given in Eq.~(\ref{eq:psi1} - \ref{eq:psi3}) respectively. In Fig.~\ref{fig:cor2}(a-b), for $U=0$, $\Gamma_{ij}$ (calculated at time $t=7J^{-1}$) shows four peaks due to the fact that the particle wave functions spread equal distance in both directions from the initial position. With the increase in $U$, the elements in the upper triangle along with the diagonal elements of the $\Gamma_{ij}$ matrix start to decrease and eventually vanish for large enough $U$. This is because the two particles avoid each other due to strong repulsion. When the two particles start from the edges Eq.~\ref{eq:psi3}, the correlation matrix $\Gamma_{ij}$ behaves differently compared to the other two cases. In Fig.~\ref{fig:cor2}(c), we plot $\Gamma_{ij}$ at time $t=20J^{-1}$, for which the corresponding local densities of the individual components $\langle n_\sigma \rangle$ are shown in Fig.~\ref{fig:ni5}(d). Since for vanishing $U$ the wave functions transmit through each other and travel to the opposite directions, we see only one peak in the correlation matrix. However, for strong enough interaction ($U>10$), the peak in the correlation matrix flips to a different position because of strong repulsion between the particles which is also visible from Fig.~\ref{fig:ni3}(c). Note that there is no doublon formation in these cases.

\subsubsection*{Effect of distance}
\begin{figure}[t]
	%\hspace{2mm}
	\includegraphics[width=0.3\columnwidth]{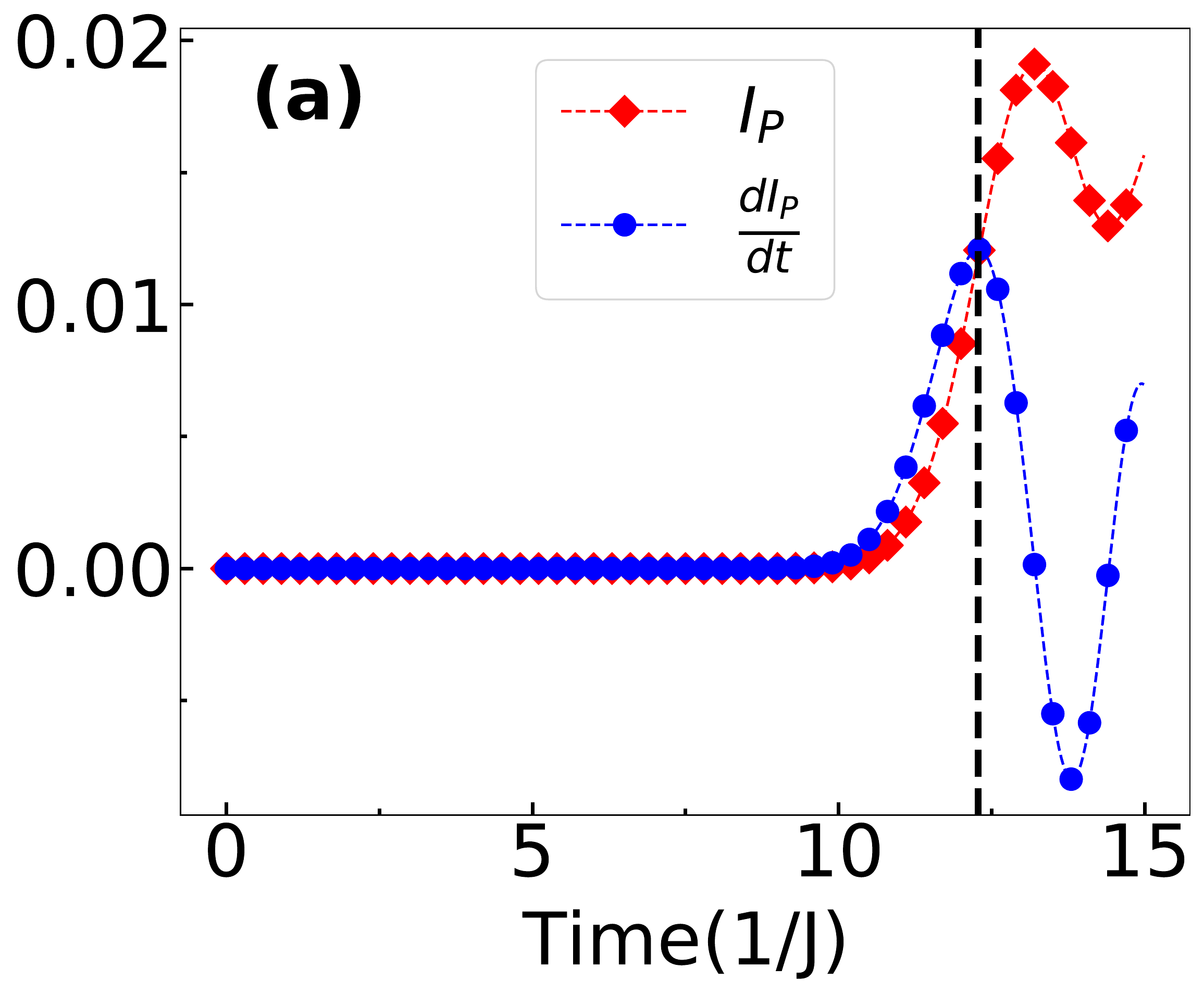}
	\hspace{2mm}
	\includegraphics[width=0.34\columnwidth]{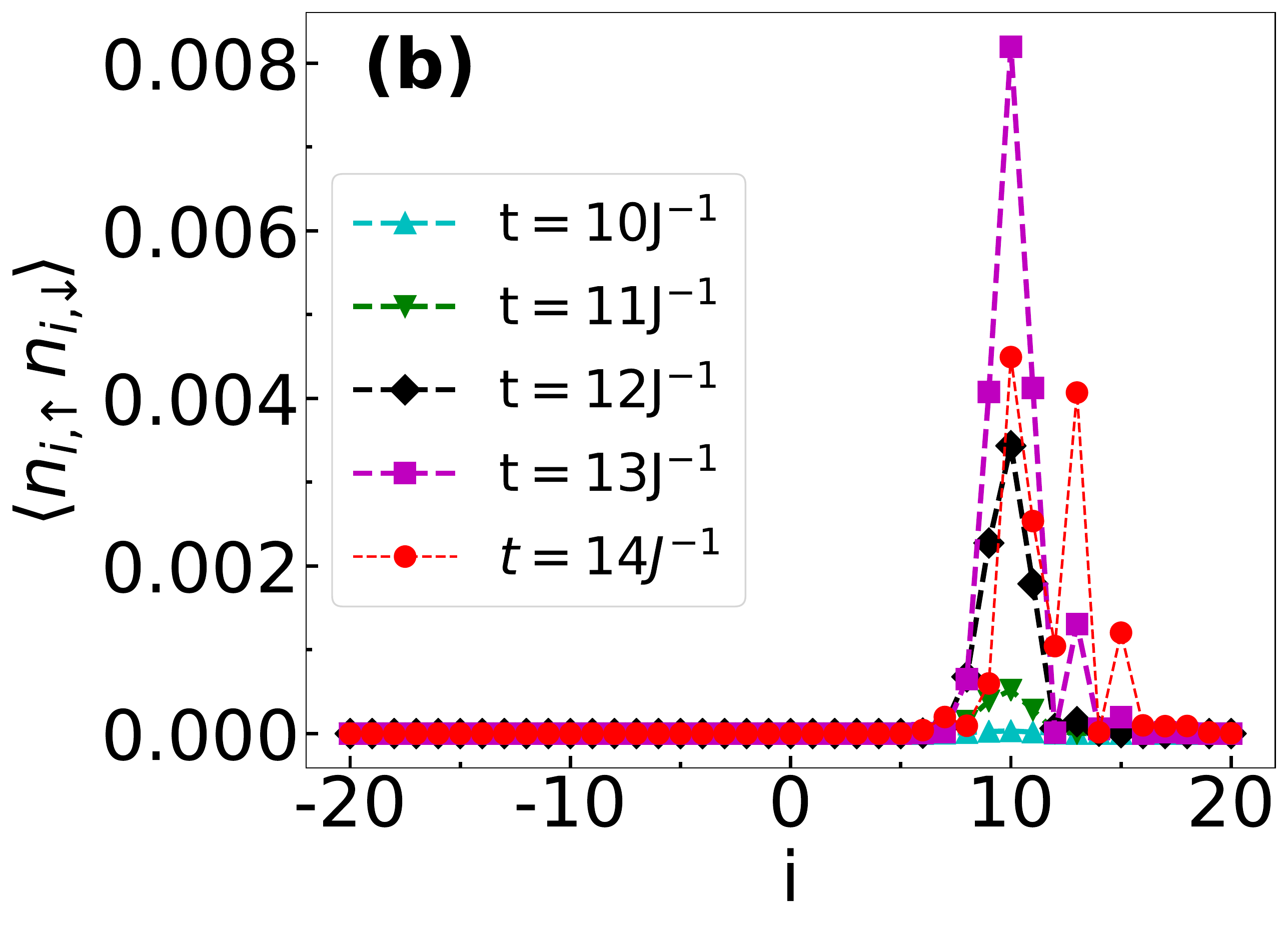}
	\hspace{2mm}
	\includegraphics[width=0.3\columnwidth]{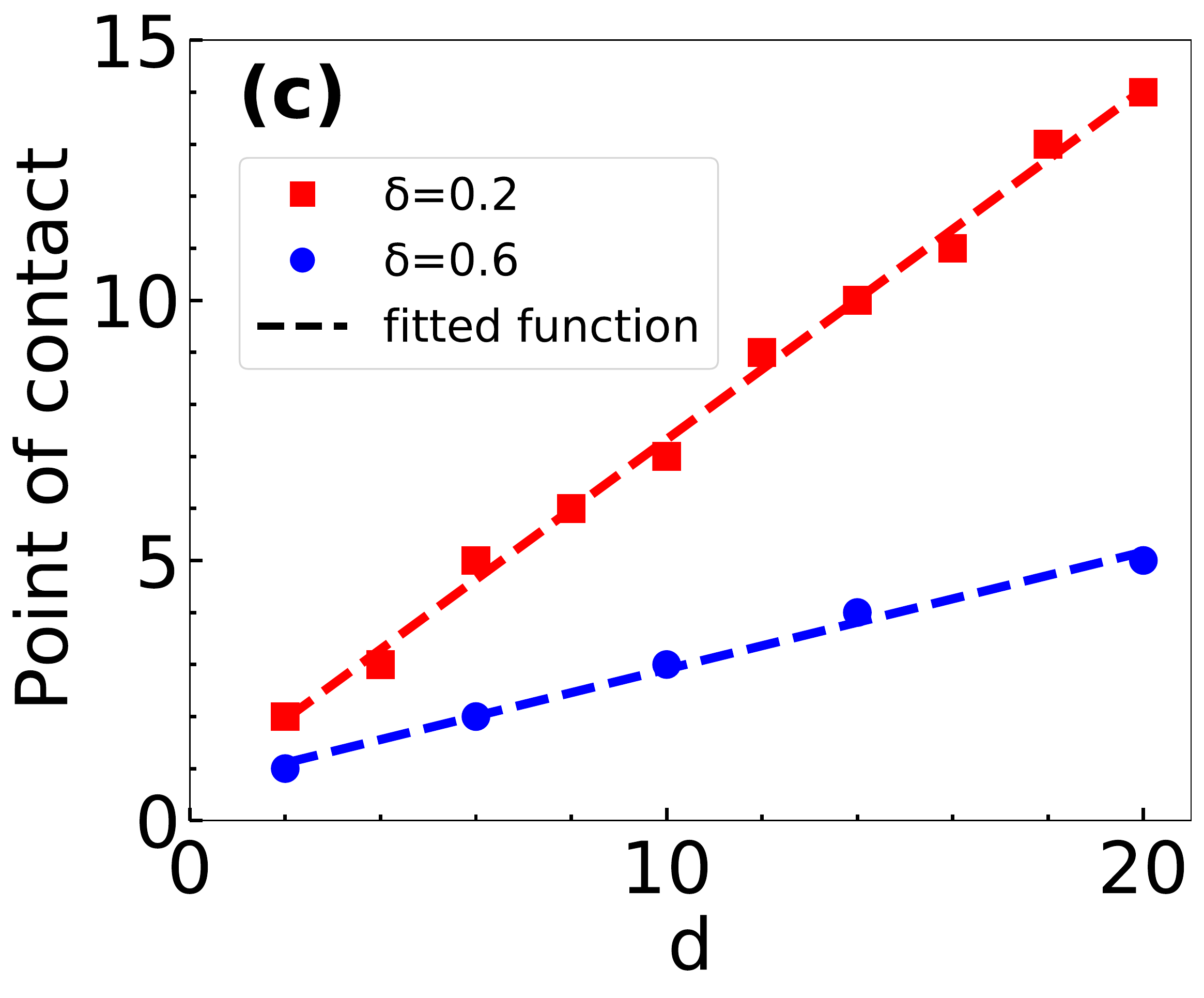}
	%\end{center}
	%\vspace*{-2mm}
	\caption{(a) $I_{P}$ and $\frac{dI_{P}}{dt}$ are plotted with respect to time. Black dashed line represents the time at which $\frac{dI_{P}}{dt}$ is maximum. Here we consider the initial state $|\Psi_0\rangle=a_{-14,\ua}^\dagger a_{14,\da}^\dagger|vac\rangle$ and  $\delta=0.2$. (b) The point of contact is shown by plotting $\left\langle n_{i,\ua}n_{i,\da}\right\rangle $ with respect to the site at different times. The time  $t=12.28J^{-1}$ corresponds to the black dashed line of (a). (c) Point of contact of the two particle wave functions are plotted with different $d$ of the initial state $|\Psi_0\rangle=a_{-d,\ua}^\dagger a_{d,\da}^\dagger|vac\rangle$. The red squares and blue circles are the data for $\delta=0.2$ and $0.6$ respectively. The dashed lines are the fitted functions.}
	\label{fig:diag_cor}
\end{figure}

From the above discussion, it is understood that the features in the time evolution of densities in the presence of \mrinal{hopping} imbalance and interaction have a strong dependence on the initial states. The point of contact of the two particle wave function strongly depends on the distance between the particles. In order to examine this we study the effect of distance between the two particles at the initial position on the QW by defining a general initial state  
\begin{equation}\label{eq:psi_d}
 |\Psi(0)\rangle=a_{-d,\ua}^\dagger a_{d,\da}^\dagger|vac\rangle,
\end{equation}
where $d$ is the distance of the occupied sites from the central one. The point at which the two particles first meet can be computed by tracking the position where the occupancy of both the \mrinal{$\ua$} and \mrinal{$\da$} particles becomes finite in the entire lattice for the first time during the time evolution. For this purpose we define a quantity 
\begin{equation}
 I_P  =  \sum_i \langle n_{i,\ua}n_{i,\da}\rangle,
\end{equation}
which becomes finite only when any site will have finite densities of both the components during the time evolution. The time evolution of $I_P$ (red squares) for an exemplary initial state $|\Psi_0\rangle=a_{-14,\ua}^\dagger a_{14,\da}^\dagger|vac\rangle$ of non interacting particles ($U=0$) and $\delta=0.2$ is shown in Fig.~\ref{fig:diag_cor}(a). This clearly shows that $I_P$ becomes finite after a certain time of evolution indicating the point of contact between the two wave functions. The actual point of contact is not easy to estimate from the figure due to the smooth variation of $I_P$ with time. To estimate the point of contact, we first plot $dI_P/dt$ (blue circles) as a function of time and obtain the time of contact as the first peak in $dI_P/dt$ which appears at $t=12.28J^{-1}$. Then we plot $\langle n_{i,\ua}n_{i,\da}\rangle$ as a function of site index $i$ for different $t$ around $t=12.28J^{-1}$ such as $t=10J^{-1},~11J^{-1},~12J^{-1},~13J^{-1} ~\rm{and} ~14J^{-1}$ in Fig.~\ref{fig:diag_cor}(b). The appearance of large peaks at $i=10$ for $t\geq 12$ is a clear indication of the point of contact. We repeat this procedure for different values of $d$ and plot the point of contact as a function of $d$ in Fig.~\ref{fig:diag_cor}(c) for two different values of $\delta$. These curves exhibit linear behaviour which can be attributed to the ballistic nature of the QW. Moreover, we find that the slopes of the fitted functions decrease with an increase in $\delta$. It is to be noted that the point of contact for all $d$ and $\delta$ is independent of $U$ as expected. However, the dependence of $U$ on $d$ can only be realized after the point of contact which will be discussed in the following subsection.  

%%%%%%%%%%%%%%%%%%%%%%%%%%%%%%%%%%%%%%%%%%%%%%%%%%%%%%%%%%%%%%%%%%%%%%%%%%%%%%%%%%%%%%%%%%%%%%%%%%%%%%
%\begin{figure}[b]
%	\begin{center}
%		\includegraphics[width=1.0\columnwidth]{Transmission_co-effecient_critical_points_t_07_t_23.pdf}
%	\end{center}
%	\caption{The interspecies interaction $U$ at which transmission co-efficient becomes zero (T=0) is plotted for different $\delta$ by evolving the initial state Eq.~\ref{eq:psi1} to  $t = 7J^{-1}$ (red star points), Eq.~\ref{eq:psi2} to $t = 7J^{-1}$ (blue squared points) and Eq.~\ref{eq:psi3} to $t = 23J^{-1}$ (magenta diamond points).}
%	\label{fig:critical transmission plot}
%\end{figure}
%%%%%%%%%%%%%%%%%%%%%%%%%%%%%%%%%%%%%%%%%%%%%%%%%%%%%%%%%%%%%%%%%%%%%%%%%%%%%%%%%%%%%%%%%%%%%%%%%%%%%%
%%%%%%%%%%%%%%%%%%%%%%%%%%%%%%%%%%%%%%%%%%%%%%%%%%
%\begin{figure}[t]
%	\begin{center}
%		\includegraphics[width=0.45\columnwidth]{Point_of_impact_Plot_vs_distance.pdf}
%	\end{center}
%	%\vspace*{-2mm}
%	\caption{Point of contact of the two particle wavefunctions (with respect to the central site) are plotted with different $d$ of the initial state 
%		$|\Psi_0\rangle=a_{-d,\ua}^\dagger a_{d,\da}^\dagger|vac\rangle$. The red squares and blue circles are the data for $\delta=0.2$ and $0.6$ respectively. The dashed lines are the fitted functions. }
%	\label{fig:position}
%\end{figure}
%%%%%%%%%%%%%%%%%%%%%%%%%%%%%%%%%%%%%%%%%%%%%%%%%%
\subsubsection*{Transmission coefficient}
%%%%%%%%%%%%%%%%%%%%%%%%%%%%%%%%%%%%%%%%%%%%%%%%%%%%%%%%%%%%%%%%%%%%%%%%%%
\begin{figure}[!t]
	%\begin{center}
	\includegraphics[width=0.338\columnwidth]{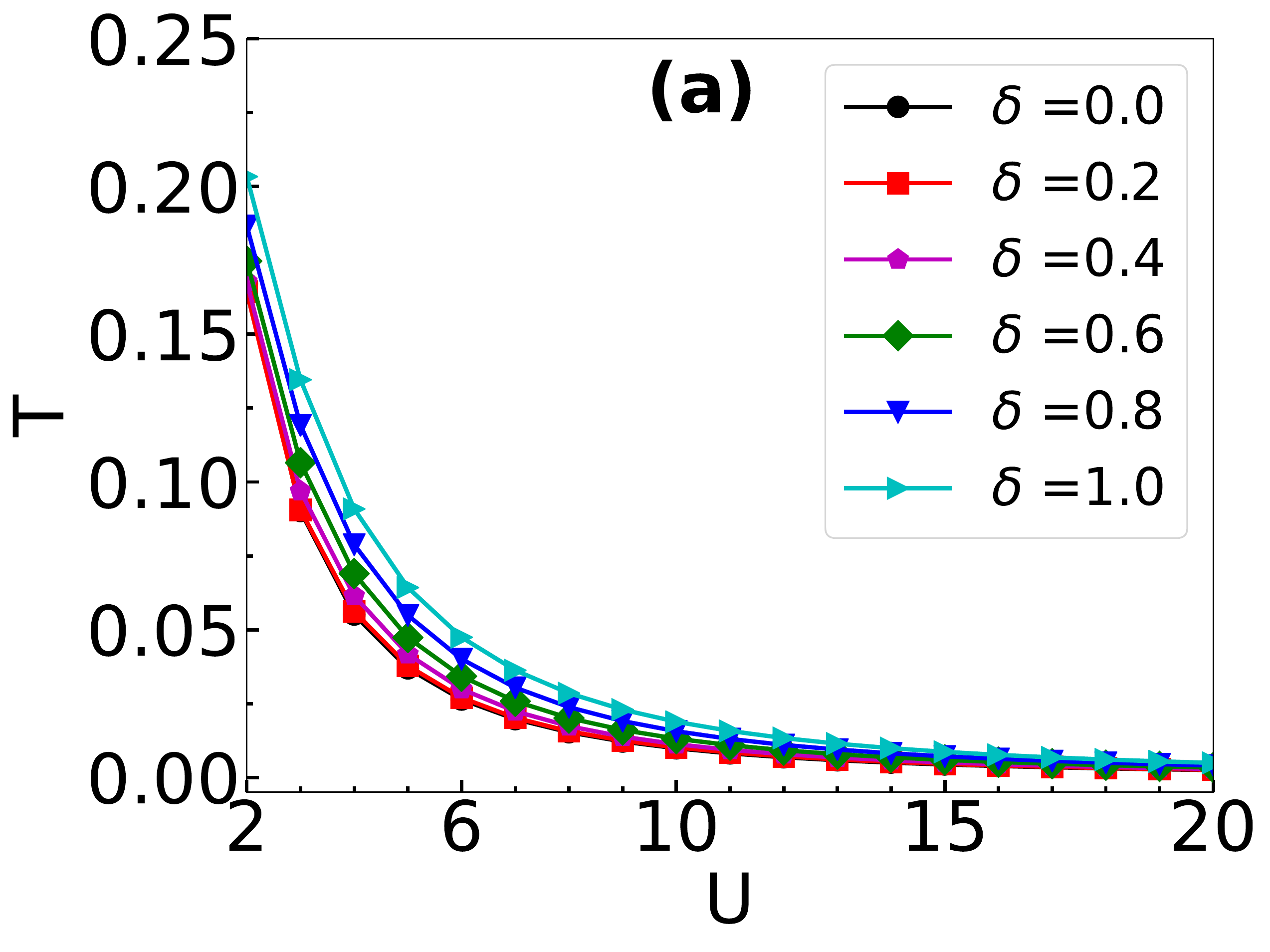}
	%\hspace*{2mm}
	\includegraphics[width=0.32\columnwidth]{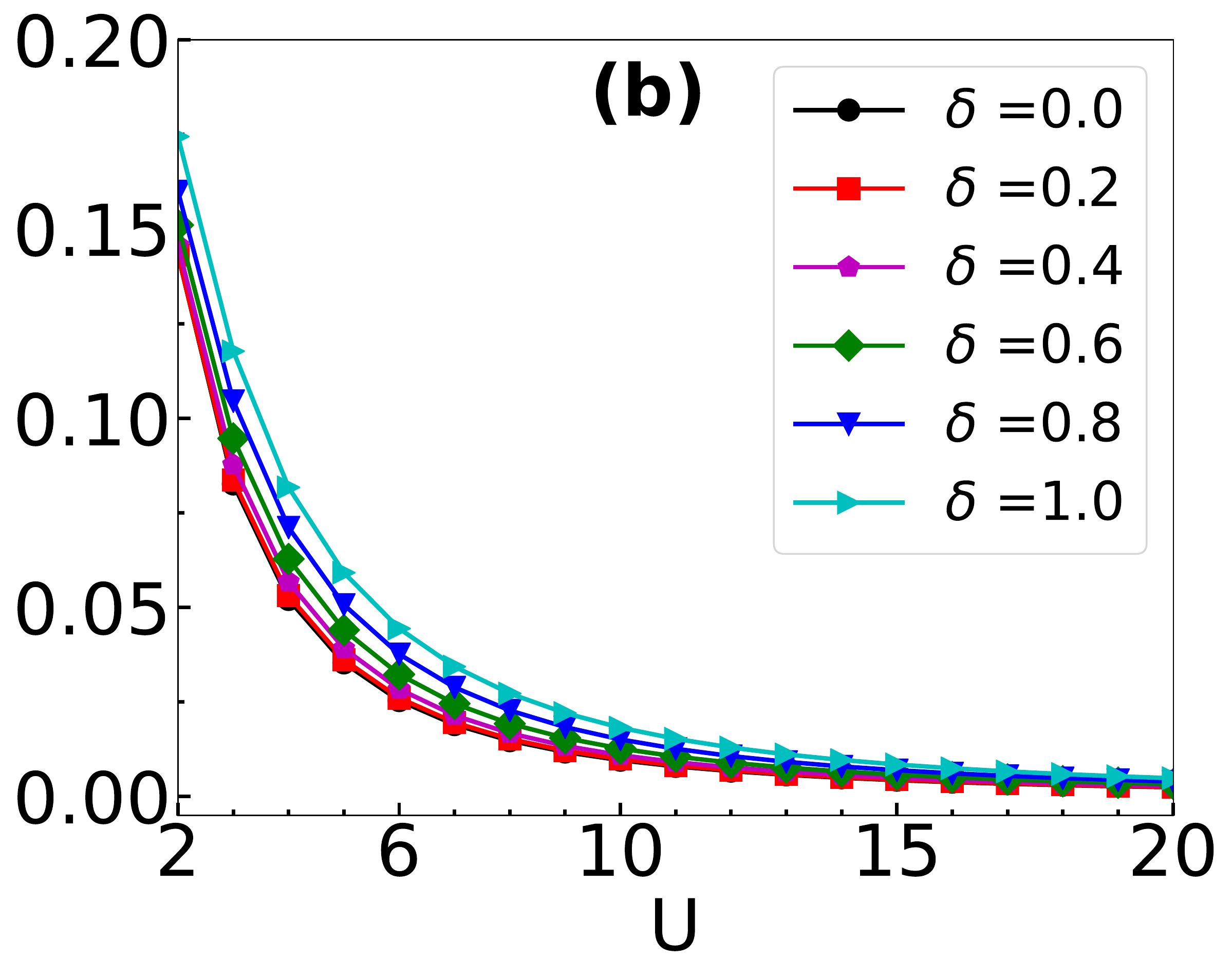}
	%\centering
	%\hspace*{2mm}
	\includegraphics[width=0.305\columnwidth]{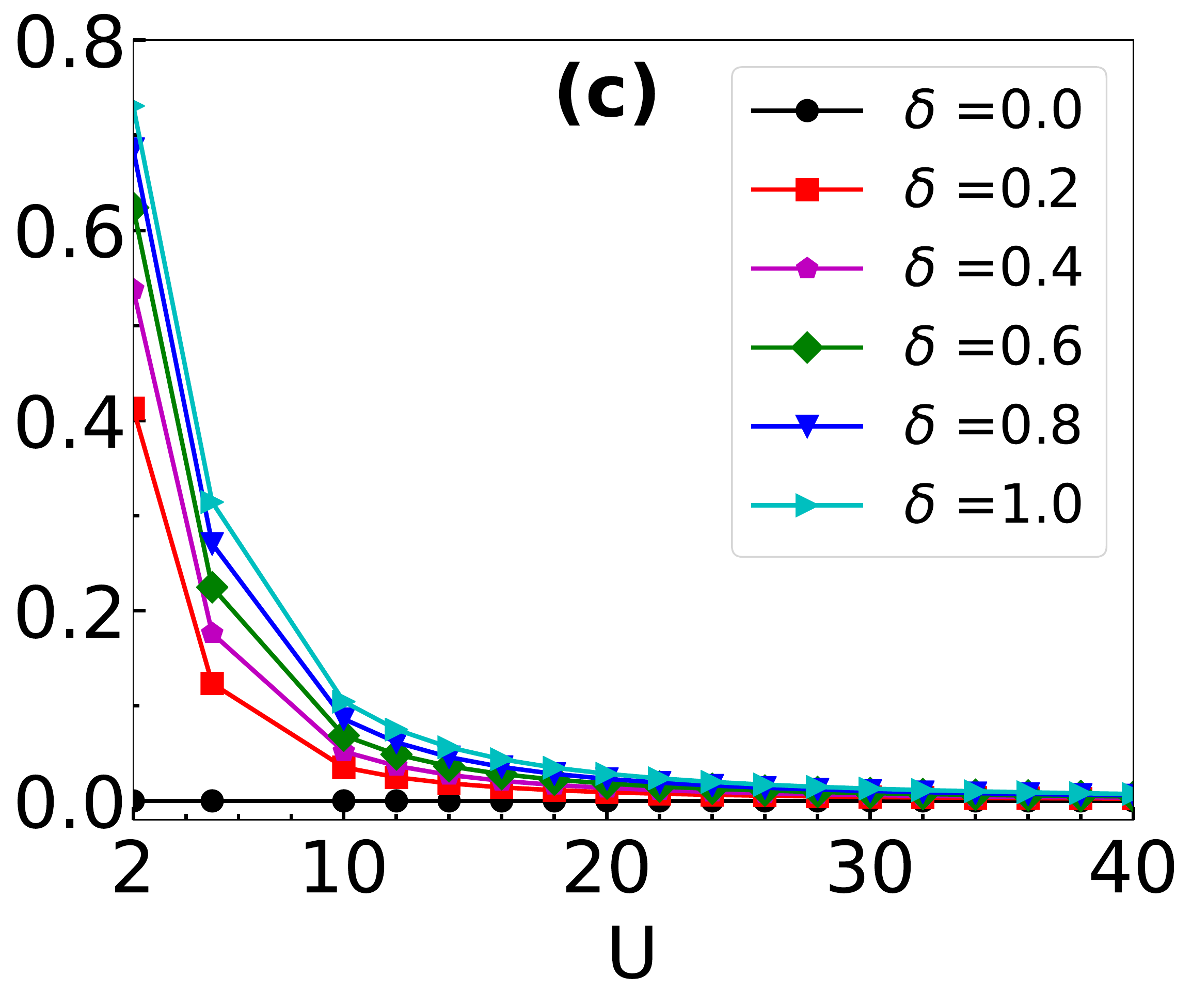}
	%\end{center}
	%\vspace*{-3mm}
	\caption{The transmission coefficient $T$ is plotted with respect to $U$ for different values of $\delta$ by evolving the initial state $|\Psi(0)\rangle$ to (a) $t = 7J^{-1}$, (b)$t = 7J^{-1}$ and $t = 23J^{-1}$. (a), (b) and (c) correspond to the results obtained using the initial states given in Eq.~\ref{eq:psi1}, Eq.~\ref{eq:psi2} and Eq.~\ref{eq:psi3} respectively.}
	\label{fig:Tnn}
\end{figure}

\begin{figure}[!t]
	%\begin{center}
	\includegraphics[width=0.45\columnwidth]{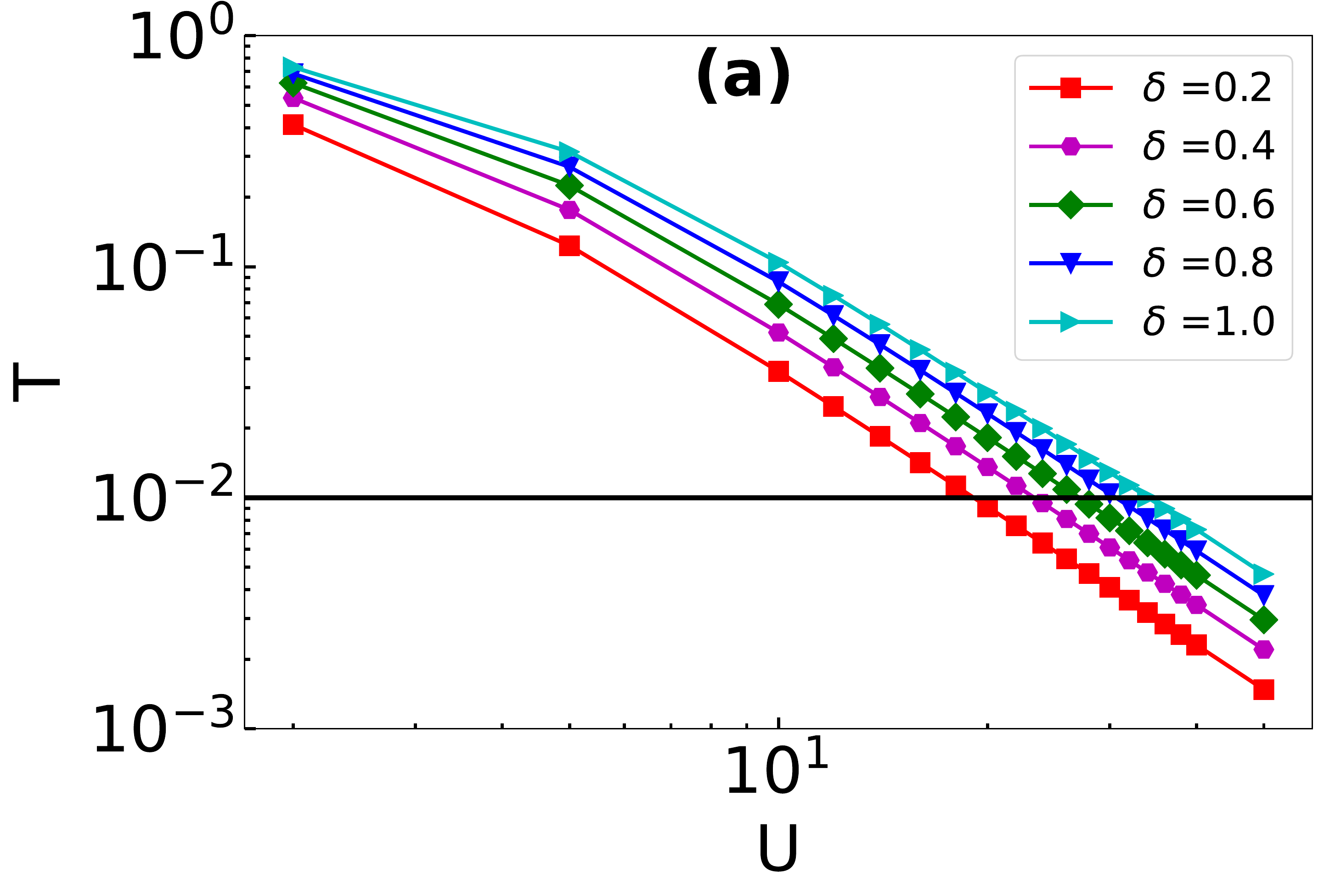}
	\hspace*{4mm}
	\includegraphics[width=0.43\columnwidth]{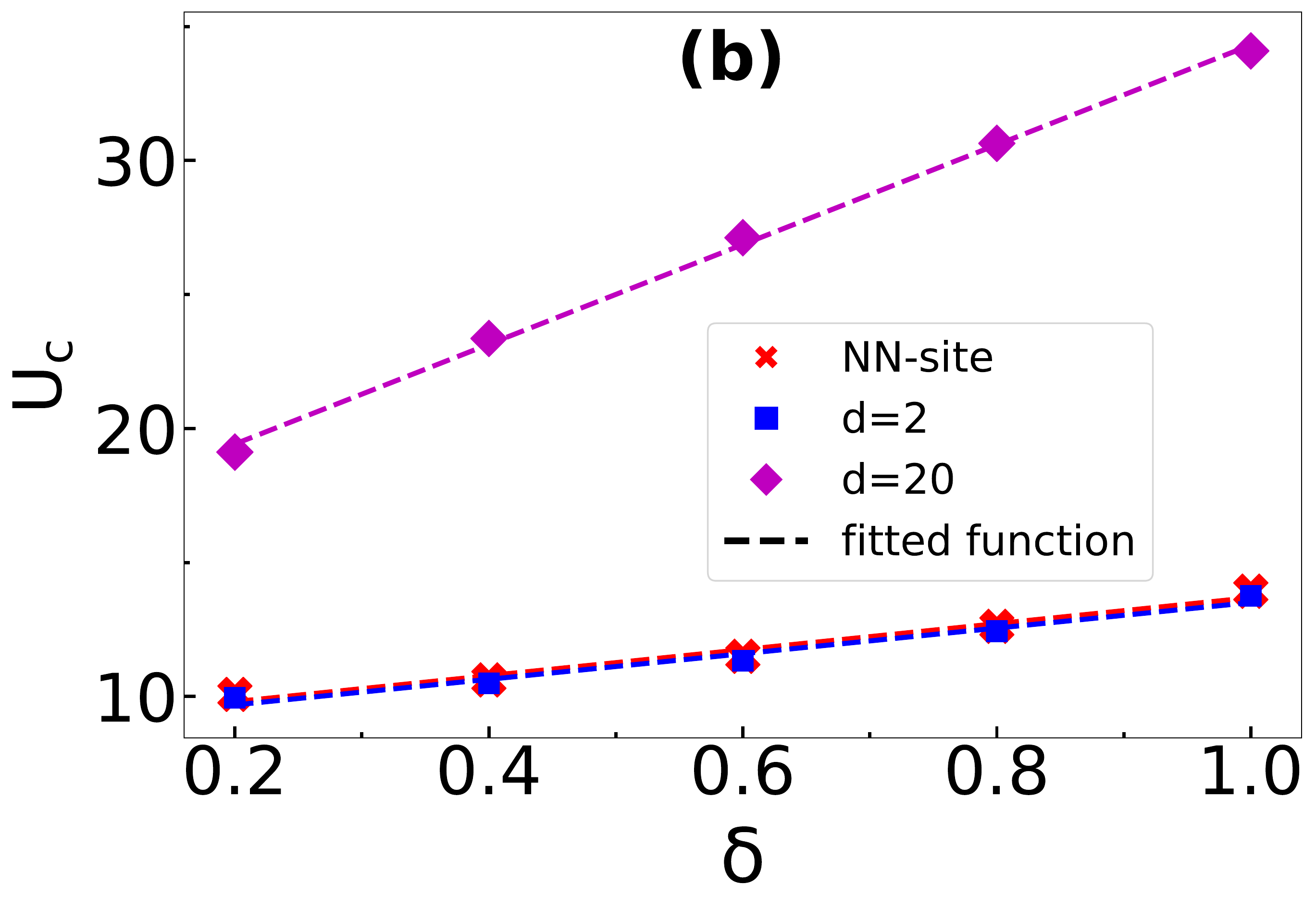}
	%\end{center}
	%\vspace*{-2mm}
	\caption{(a) The transmission coefficient $T$ is plotted as a function of $U$ for different values of $\delta$ in the log-log scale for the initial state given in Eq.~\ref{eq:psi3} at $t = 23J^{-1}$. The dashed line marks $T=10^{-2}$ which is considered as the critical $T$ for zero transmission and its point of intersection with different curves are the corresponding $U_c$. (b) The plot between $U_c$ and $\delta$ is obtained by evolving the initial states given in Eq.~\ref{eq:psi1} to  $t = 7J^{-1}$ (red stars), Eq.~\ref{eq:psi2} to $t = 7J^{-1}$ (blue squares) and Eq.~\ref{eq:psi3} to $t = 23J^{-1}$ (magenta diamonds).}
	\label{fig:log-log plot}
\end{figure}

The effect of hopping imbalance on the QW is further studied by calculating the transmission coefficient defined as 
\begin{equation}
 T = \sum\limits_{\substack{i,j \\ j>i}} \langle n_{i,\ua} n_{j,\da}\rangle.
\end{equation}
which is nothing but the sum over all the upper triangular elements of the correlation matrix. This provides an estimate of the probability of the existence of the $\ua$ particle on the right side region of the $\da$ particle profile at a particular instant during the QW. In order to understand the behaviour of $T$ of interacting particles with hopping imbalance, we plot $T$ with respect to $U$ for different values of $\delta$ in Fig.~\ref{fig:Tnn}(a), (b) and (c) for three initial states given in Eq.~\ref{eq:psi1}, Eq.~\ref{eq:psi2} and Eq.~\ref{eq:psi3} respectively. In all these cases we observe that the values of $T$ decrease with increase in $U$ and gradually vanish in the limit of strong interactions. Moreover, a larger hopping imbalance (i.e. smaller $\delta$) leads to a faster decay of $T$. This indicates that for a large (small) imbalance, the transmission ceases for a weak (strong) interaction $U$. This is because for small $\delta$ the on-site density of the $\da$ particle at the point of contact during the QW is larger compared to the case of larger $\delta$. Hence, at the point of contact the effective interaction experienced by the $\ua$ particle is stronger for smaller $\delta$. Note that in Fig.~\ref{fig:Tnn}(c) for $\delta=0$, the $T$ is always zero because the $\da$ particle is localized at the edge (as $J_{\da} = 0$), and the $\ua$ particle can never go past the edge due to the open boundary condition. It can be seen from Fig.~\ref{fig:Tnn} that the vanishing up of $T$ is very slow as a function of $U$ for all the cases considered. In order to obtain the value of critical interaction strength ($U_c$) for no transmission or complete reflection, we have re-plotted the $T-U$ plot in the log-log scale (see Fig.~\ref{fig:log-log plot}(a)) and estimated $U_c$ by assuming $T = 10^{-2}$ as the condition for no transmission. Using the above method, we have calculated the values of $U_c$ for different initial states and plotted them in the $U_c-\delta$ plane in Fig.~\ref{fig:log-log plot}(b). The curves for different initial states exhibit the linear dependence of $U_c$ with respect to $\delta$. Moreover, we observe that the critical strength and slope of the curves increase with increasing $d$. Note that in our analysis, we consider $U> \max{(J_{\ua},J_{\da})}$ to see the effect of $U$ on $T$.
%These clearly 

%%%%%%%%%%%%%%%%%%%%%%%%%%%%%%%%%%%%%%%%%%%%%%%%%%%%%%%%%%%%%%%%%%%%%%%%%%%%%%%%%%%%%%%%%%%%%%%%%%%%%%

%%%%%%%%%
\section*{Conclusions} 
\label{sec:co}
We have studied the QW of a two-component system in the presence of interaction and hopping imbalance in a one-dimensional lattice. By considering different initial states depending on the positions of the particles ($\ua$ and $\da$  where the $\ua$ particle has higher hopping strength), 
we have analyzed the combined effect of hopping imbalance and inter-component interaction on the two particle QW. 
We have found that when the two particles initially start from the central site of the lattice, 
the QW exhibits independent particle QWs to a QW of composite particles or doublon as a 
function of repulsive interactions. 
However, for the initial state with two particles at two different sites (a few sites apart), 
the $\ua$ particle wave function gets reflected 
from the $\da$ particle's wave function for large enough interactions, and no doublon is formed. 
On the other hand, when the two particles start from the opposite ends of the lattice, 
the situation is completely different for strong interactions. In this case, both the $\ua$ and $\da$ particle wave functions significantly
reflect from each other at a point close to the initial position of the $\da$  particle. 
While we obtain different behavior compared to the many-body limit depending upon the parameters of the model Hamiltonian, the phenomenon of zero transmission in the limit of large inter-component interaction resemble the phase separation which has been predicted in systems of binary atomic mixtures in optical lattices~\cite{mishraps,Sowi_ski_2019,PhysRevB.78.184513}. Moreover, we have obtained that the change in the initial position of the particles leads to a qualitative change in the results. These findings provide insights into the dynamical behavior of a mixture of two component systems in periodic potential at the few particle levels.  
Due to the recent experimental progress in controlled creation and manipulation of multi-component atomic mixtures in an optical lattice and the single site addressing techniques, our prediction can, in principle, be simulated in quantum gas experiments. While the hopping imbalance can be indirectly obtained by considering a two component atomic mixture of different masses such as $^{87}$Rb and $^{41}$K atoms~\cite{Catani}, it will be impossible to tune the hopping imbalance to explore the physics in broader parameter space. Therefore, an appropriate platform can be the mixture of two hyperfine states of a particular atomic species in a state dependent optical lattice where the hopping strengths of each internal state can be independently tuned~\cite{Altman2003,Duan_altman2003,Mandel_et_al2003,Soltan-Panahi2011,Jian-wei-2017}.

% Note: During the preparation of this paper, we became aware of a recent 
% work on Bloch oscillation of two-component QWers on a 
% tilted lattice~\cite{Sowinski_sarkar}. The work of Ref.\ref{Sowinski_sarkar} is focused on 
% the BO of two indistinguishable non-interacting particles of one component interacting with a third particle of another component. 
% They also considered the effect of particle statistics and interaction in the BO.  
% Another interesting situation would be if instead of starting with an initial state like
% $a^{\dagger2}_0|MI1\rangle$ we separate the QWers by one or two lattice site and study them. 
% Depending on the three-body interaction we can see how three body
% interaction influence the Hanbury Brown\UTF{2013}Twiss (HBT) interference
% ~\cite{Bromberg2009,Peruzzo2010,Sansoni2012,Broome2013,Spring2013,Tillmann2013,Crespi2013}
% and Fermionization~\cite{Lahini_walk,Greiner_walk}.
\bibliography{references}

\begin{thebibliography}{10}
\urlstyle{rm}
\expandafter\ifx\csname url\endcsname\relax
  \def\url#1{\texttt{#1}}\fi
\expandafter\ifx\csname urlprefix\endcsname\relax\def\urlprefix{URL }\fi
\expandafter\ifx\csname doiprefix\endcsname\relax\def\doiprefix{DOI: }\fi
\providecommand{\bibinfo}[2]{#2}
\providecommand{\eprint}[2][]{\url{#2}}

\bibitem{Aharonov}
\bibinfo{author}{Aharonov, Y.}, \bibinfo{author}{Davidovich, L.} \&
  \bibinfo{author}{Zagury, N.}
\newblock \bibinfo{journal}{\bibinfo{title}{Quantum random walks}}.
\newblock {\emph{\JournalTitle{Phys. Rev. A}}} \textbf{\bibinfo{volume}{48}},
  \bibinfo{pages}{1687--1690}, \doiprefix\url{10.1103/PhysRevA.48.1687}
  (\bibinfo{year}{1993}).

\bibitem{Kempe2003}
\bibinfo{author}{Kempe, J.}
\newblock \bibinfo{journal}{\bibinfo{title}{Quantum random walks: An
  introductory overview}}.
\newblock {\emph{\JournalTitle{Contemporary Physics}}}
  \textbf{\bibinfo{volume}{44}}, \bibinfo{pages}{307–327},
  \doiprefix\url{10.1080/00107151031000110776} (\bibinfo{year}{2003}).

\bibitem{Venegas-Andraca2012}
\bibinfo{author}{Venegas-Andraca, S.~E.}
\newblock \bibinfo{journal}{\bibinfo{title}{Quantum walks: a comprehensive
  review}}.
\newblock {\emph{\JournalTitle{Quantum Information Processing}}}
  \textbf{\bibinfo{volume}{11}}, \bibinfo{pages}{1015--1106},
  \doiprefix\url{10.1007/s11128-012-0432-5} (\bibinfo{year}{2012}).

\bibitem{ambainis2003quantum}
\bibinfo{author}{Ambainis, A.}
\newblock \bibinfo{journal}{\bibinfo{title}{Quantum walks and their algorithmic
  applications}}.
\newblock {\emph{\JournalTitle{International Journal of Quantum Information}}}
  \textbf{\bibinfo{volume}{1}}, \bibinfo{pages}{507--518}
  (\bibinfo{year}{2003}).

\bibitem{Kempe_application}
\bibinfo{author}{Shenvi, N.}, \bibinfo{author}{Kempe, J.} \&
  \bibinfo{author}{Whaley, K.~B.}
\newblock \bibinfo{journal}{\bibinfo{title}{Quantum random-walk search
  algorithm}}.
\newblock {\emph{\JournalTitle{Phys. Rev. A}}} \textbf{\bibinfo{volume}{67}},
  \bibinfo{pages}{052307}, \doiprefix\url{10.1103/PhysRevA.67.052307}
  (\bibinfo{year}{2003}).

\bibitem{Childs_walk}
\bibinfo{author}{Childs, A.~M.} \& \bibinfo{author}{Goldstone, J.}
\newblock \bibinfo{journal}{\bibinfo{title}{Spatial search by quantum walk}}.
\newblock {\emph{\JournalTitle{Phys. Rev. A}}} \textbf{\bibinfo{volume}{70}},
  \bibinfo{pages}{022314}, \doiprefix\url{10.1103/PhysRevA.70.022314}
  (\bibinfo{year}{2004}).

\bibitem{childs2013universal}
\bibinfo{author}{Childs, A.~M.}, \bibinfo{author}{Gosset, D.} \&
  \bibinfo{author}{Webb, Z.}
\newblock \bibinfo{journal}{\bibinfo{title}{Universal computation by
  multiparticle quantum walk}}.
\newblock {\emph{\JournalTitle{Science}}} \textbf{\bibinfo{volume}{339}},
  \bibinfo{pages}{791--794} (\bibinfo{year}{2013}).

\bibitem{Schuld_2014}
\bibinfo{author}{Schuld, M.}, \bibinfo{author}{Sinayskiy, I.} \&
  \bibinfo{author}{Petruccione, F.}
\newblock \bibinfo{journal}{\bibinfo{title}{The quest for a quantum neural
  network}}.
\newblock {\emph{\JournalTitle{Quantum Information Processing}}}
  \textbf{\bibinfo{volume}{13}}, \bibinfo{pages}{2567–2586},
  \doiprefix\url{10.1007/s11128-014-0809-8} (\bibinfo{year}{2014}).

\bibitem{Ryan_QW}
\bibinfo{author}{Ryan, C.~A.}, \bibinfo{author}{Laforest, M.},
  \bibinfo{author}{Boileau, J.~C.} \& \bibinfo{author}{Laflamme, R.}
\newblock \bibinfo{journal}{\bibinfo{title}{Experimental implementation of a
  discrete-time quantum random walk on an nmr quantum-information processor}}.
\newblock {\emph{\JournalTitle{Phys. Rev. A}}} \textbf{\bibinfo{volume}{72}},
  \bibinfo{pages}{062317}, \doiprefix\url{10.1103/PhysRevA.72.062317}
  (\bibinfo{year}{2005}).

\bibitem{Schmitz2009}
\bibinfo{author}{Schmitz, H.} \emph{et~al.}
\newblock \bibinfo{journal}{\bibinfo{title}{Quantum walk of a trapped ion in
  phase space}}.
\newblock {\emph{\JournalTitle{Phys. Rev. Lett.}}}
  \textbf{\bibinfo{volume}{103}}, \bibinfo{pages}{090504},
  \doiprefix\url{10.1103/PhysRevLett.103.090504} (\bibinfo{year}{2009}).

\bibitem{Zahringer2010}
\bibinfo{author}{Z\"ahringer, F.} \emph{et~al.}
\newblock \bibinfo{journal}{\bibinfo{title}{Realization of a quantum walk with
  one and two trapped ions}}.
\newblock {\emph{\JournalTitle{Phys. Rev. Lett.}}}
  \textbf{\bibinfo{volume}{104}}, \bibinfo{pages}{100503},
  \doiprefix\url{10.1103/PhysRevLett.104.100503} (\bibinfo{year}{2010}).

\bibitem{Karski2009}
\bibinfo{author}{Karski, M.} \emph{et~al.}
\newblock \bibinfo{journal}{\bibinfo{title}{Quantum walk in position space with
  single optically trapped atoms}}.
\newblock {\emph{\JournalTitle{Science}}} \textbf{\bibinfo{volume}{325}},
  \bibinfo{pages}{174--177}, \doiprefix\url{10.1126/science.1174436}
  (\bibinfo{year}{2009}).

\bibitem{Weitenberg2011}
\bibinfo{author}{Weitenberg, C.} \emph{et~al.}
\newblock \bibinfo{journal}{\bibinfo{title}{Single-spin addressing in an atomic
  mott insulator}}.
\newblock {\emph{\JournalTitle{Nature}}} \textbf{\bibinfo{volume}{471}},
  \bibinfo{pages}{319--324} (\bibinfo{year}{2011}).

\bibitem{Fukuhara2013}
\bibinfo{author}{Fukuhara, T.} \emph{et~al.}
\newblock \bibinfo{journal}{\bibinfo{title}{Microscopic observation of magnon
  bound states and their dynamics}}.
\newblock {\emph{\JournalTitle{Nature}}} \textbf{\bibinfo{volume}{502}},
  \bibinfo{pages}{76 EP --} (\bibinfo{year}{2013}).

\bibitem{Manouchehri2014}
\bibinfo{author}{Manouchehri, K.} \& \bibinfo{author}{Wang, J.}
\newblock \bibinfo{journal}{\bibinfo{title}{Physical implementation of quantum
  walks}}.
\newblock {\emph{\JournalTitle{Physical Implementation of Quantum Walks,
  Springer}}} \doiprefix\url{10.1007/978-3-642-36014-5} (\bibinfo{year}{2014}).

\bibitem{Hoyer2010}
\bibinfo{author}{Hoyer, S.}, \bibinfo{author}{Sarovar, M.} \&
  \bibinfo{author}{Whaley, K.~B.}
\newblock \bibinfo{journal}{\bibinfo{title}{Limits of quantum speedup in
  photosynthetic light harvesting}}.
\newblock {\emph{\JournalTitle{New Journal of Physics}}}
  \textbf{\bibinfo{volume}{12}}, \bibinfo{pages}{065041},
  \doiprefix\url{10.1088/1367-2630/12/6/065041} (\bibinfo{year}{2010}).

\bibitem{Mohseni2008}
\bibinfo{author}{Mohseni, M.}, \bibinfo{author}{Rebentrost, P.},
  \bibinfo{author}{Lloyd, S.} \& \bibinfo{author}{Aspuru-Guzik, A.}
\newblock \bibinfo{journal}{\bibinfo{title}{Environment-assisted quantum walks
  in photosynthetic energy transfer}}.
\newblock {\emph{\JournalTitle{The Journal of Chemical Physics}}}
  \textbf{\bibinfo{volume}{129}}, \bibinfo{pages}{174106},
  \doiprefix\url{10.1063/1.3002335} (\bibinfo{year}{2008}).

\bibitem{PhysRevLett.119.130501}
\bibinfo{author}{Zhan, X.} \emph{et~al.}
\newblock \bibinfo{journal}{\bibinfo{title}{Detecting topological invariants in
  nonunitary discrete-time quantum walks}}.
\newblock {\emph{\JournalTitle{Phys. Rev. Lett.}}}
  \textbf{\bibinfo{volume}{119}}, \bibinfo{pages}{130501},
  \doiprefix\url{10.1103/PhysRevLett.119.130501} (\bibinfo{year}{2017}).

\bibitem{PhysRevX.7.031023}
\bibinfo{author}{Flurin, E.} \emph{et~al.}
\newblock \bibinfo{journal}{\bibinfo{title}{Observing topological invariants
  using quantum walks in superconducting circuits}}.
\newblock {\emph{\JournalTitle{Phys. Rev. X}}} \textbf{\bibinfo{volume}{7}},
  \bibinfo{pages}{031023}, \doiprefix\url{10.1103/PhysRevX.7.031023}
  (\bibinfo{year}{2017}).

\bibitem{PhysRevLett.114.140502}
\bibinfo{author}{Xue, P.} \emph{et~al.}
\newblock \bibinfo{journal}{\bibinfo{title}{Experimental quantum-walk revival
  with a time-dependent coin}}.
\newblock {\emph{\JournalTitle{Phys. Rev. Lett.}}}
  \textbf{\bibinfo{volume}{114}}, \bibinfo{pages}{140502},
  \doiprefix\url{10.1103/PhysRevLett.114.140502} (\bibinfo{year}{2015}).

\bibitem{chalabi2019synthetic}
\bibinfo{author}{Chalabi, H.} \emph{et~al.}
\newblock \bibinfo{journal}{\bibinfo{title}{Synthetic gauge field for
  two-dimensional time-multiplexed quantum random walks}}.
\newblock {\emph{\JournalTitle{Physical review letters}}}
  \textbf{\bibinfo{volume}{123}}, \bibinfo{pages}{150503}
  (\bibinfo{year}{2019}).

\bibitem{schreiber2011decoherence}
\bibinfo{author}{Schreiber, A.} \emph{et~al.}
\newblock \bibinfo{journal}{\bibinfo{title}{Decoherence and disorder in quantum
  walks: from ballistic spread to localization}}.
\newblock {\emph{\JournalTitle{Physical review letters}}}
  \textbf{\bibinfo{volume}{106}}, \bibinfo{pages}{180403}
  (\bibinfo{year}{2011}).

\bibitem{PhysRevA.86.063632}
\bibinfo{author}{Killi, M.}, \bibinfo{author}{Trotzky, S.} \&
  \bibinfo{author}{Paramekanti, A.}
\newblock \bibinfo{journal}{\bibinfo{title}{Anisotropic quantum quench in the
  presence of frustration or background gauge fields: A probe of bulk currents
  and topological chiral edge modes}}.
\newblock {\emph{\JournalTitle{Phys. Rev. A}}} \textbf{\bibinfo{volume}{86}},
  \bibinfo{pages}{063632}, \doiprefix\url{10.1103/PhysRevA.86.063632}
  (\bibinfo{year}{2012}).

\bibitem{PhysRevA.101.032336}
\bibinfo{author}{Razzoli, L.}, \bibinfo{author}{Paris, M. G.~A.} \&
  \bibinfo{author}{Bordone, P.}
\newblock \bibinfo{journal}{\bibinfo{title}{Continuous-time quantum walks on
  planar lattices and the role of the magnetic field}}.
\newblock {\emph{\JournalTitle{Phys. Rev. A}}} \textbf{\bibinfo{volume}{101}},
  \bibinfo{pages}{032336}, \doiprefix\url{10.1103/PhysRevA.101.032336}
  (\bibinfo{year}{2020}).

\bibitem{Eugene_et_al}
\bibinfo{author}{Kitagawa, T.}, \bibinfo{author}{Rudner, M.~S.},
  \bibinfo{author}{Berg, E.} \& \bibinfo{author}{Demler, E.}
\newblock \bibinfo{journal}{\bibinfo{title}{Exploring topological phases with
  quantum walks}}.
\newblock {\emph{\JournalTitle{Phys. Rev. A}}} \textbf{\bibinfo{volume}{82}},
  \bibinfo{pages}{033429}, \doiprefix\url{10.1103/PhysRevA.82.033429}
  (\bibinfo{year}{2010}).

\bibitem{PhysRevB.86.195414}
\bibinfo{author}{Asb\'oth, J.~K.}
\newblock \bibinfo{journal}{\bibinfo{title}{Symmetries, topological phases, and
  bound states in the one-dimensional quantum walk}}.
\newblock {\emph{\JournalTitle{Phys. Rev. B}}} \textbf{\bibinfo{volume}{86}},
  \bibinfo{pages}{195414}, \doiprefix\url{10.1103/PhysRevB.86.195414}
  (\bibinfo{year}{2012}).

\bibitem{wu2019topological}
\bibinfo{author}{Wu, J.}, \bibinfo{author}{Zhang, W.-W.} \&
  \bibinfo{author}{Sanders, B.~C.}
\newblock \bibinfo{journal}{\bibinfo{title}{Topological quantum walks: Theory
  and experiments}}.
\newblock {\emph{\JournalTitle{Frontiers of Physics}}}
  \textbf{\bibinfo{volume}{14}}, \bibinfo{pages}{1--6} (\bibinfo{year}{2019}).

\bibitem{peng-xue2015}
\bibinfo{author}{Xue, P.} \emph{et~al.}
\newblock \bibinfo{journal}{\bibinfo{title}{Localized state in a
  two-dimensional quantum walk on a disordered lattice}}.
\newblock {\emph{\JournalTitle{Phys. Rev. A}}} \textbf{\bibinfo{volume}{92}},
  \bibinfo{pages}{042316}, \doiprefix\url{10.1103/PhysRevA.92.042316}
  (\bibinfo{year}{2015}).

\bibitem{lozada-Vera2016}
\bibinfo{author}{Lozada-Vera, J.} \emph{et~al.}
\newblock \bibinfo{journal}{\bibinfo{title}{Quantum simulation of the anderson
  hamiltonian with an array of coupled nanoresonators: delocalization and
  thermalization effects}}.
\newblock {\emph{\JournalTitle{EPJ Quantum Technology}}}
  \textbf{\bibinfo{volume}{3}}, \bibinfo{pages}{9},
  \doiprefix\url{10.1140/epjqt/s40507-016-0047-3} (\bibinfo{year}{2016}).

\bibitem{Bromberg2009}
\bibinfo{author}{Bromberg, Y.}, \bibinfo{author}{Lahini, Y.},
  \bibinfo{author}{Morandotti, R.} \& \bibinfo{author}{Silberberg, Y.}
\newblock \bibinfo{journal}{\bibinfo{title}{Quantum and classical correlations
  in waveguide lattices}}.
\newblock {\emph{\JournalTitle{Phys. Rev. Lett.}}}
  \textbf{\bibinfo{volume}{102}}, \bibinfo{pages}{253904},
  \doiprefix\url{10.1103/PhysRevLett.102.253904} (\bibinfo{year}{2009}).

\bibitem{Peruzzo2010}
\bibinfo{author}{Peruzzo, A.} \emph{et~al.}
\newblock \bibinfo{journal}{\bibinfo{title}{Quantum walks of correlated
  photons}}.
\newblock {\emph{\JournalTitle{Science}}} \textbf{\bibinfo{volume}{329}},
  \bibinfo{pages}{1500--1503}, \doiprefix\url{10.1126/science.1193515}
  (\bibinfo{year}{2010}).

\bibitem{Sansoni2012}
\bibinfo{author}{Sansoni, L.} \emph{et~al.}
\newblock \bibinfo{journal}{\bibinfo{title}{Two-particle bosonic-fermionic
  quantum walk via integrated photonics}}.
\newblock {\emph{\JournalTitle{Phys. Rev. Lett.}}}
  \textbf{\bibinfo{volume}{108}}, \bibinfo{pages}{010502},
  \doiprefix\url{10.1103/PhysRevLett.108.010502} (\bibinfo{year}{2012}).

\bibitem{Broome2013}
\bibinfo{author}{Broome, M.~A.} \emph{et~al.}
\newblock \bibinfo{journal}{\bibinfo{title}{Photonic boson sampling in a
  tunable circuit}}.
\newblock {\emph{\JournalTitle{Science}}} \textbf{\bibinfo{volume}{339}},
  \bibinfo{pages}{794--798}, \doiprefix\url{10.1126/science.1231440}
  (\bibinfo{year}{2013}).

\bibitem{Spring2013}
\bibinfo{author}{Spring, J.~B.} \emph{et~al.}
\newblock \bibinfo{journal}{\bibinfo{title}{Boson sampling on a photonic
  chip}}.
\newblock {\emph{\JournalTitle{Science}}} \textbf{\bibinfo{volume}{339}},
  \bibinfo{pages}{798--801}, \doiprefix\url{10.1126/science.1231692}
  (\bibinfo{year}{2013}).

\bibitem{Tillmann2013}
\bibinfo{author}{Tillmann, M.} \emph{et~al.}
\newblock \bibinfo{journal}{\bibinfo{title}{Experimental boson sampling}}.
\newblock {\emph{\JournalTitle{Nature Photonics}}}
  \textbf{\bibinfo{volume}{7}}, \bibinfo{pages}{540 EP --}
  (\bibinfo{year}{2013}).

\bibitem{Crespi2013}
\bibinfo{author}{Crespi, A.} \emph{et~al.}
\newblock \bibinfo{journal}{\bibinfo{title}{Integrated multimode
  interferometers with arbitrary designs for photonic boson sampling}}.
\newblock {\emph{\JournalTitle{Nature Photonics}}}
  \textbf{\bibinfo{volume}{7}}, \bibinfo{pages}{545 EP --}
  (\bibinfo{year}{2013}).

\bibitem{Greiner_walk}
\bibinfo{author}{Preiss, P.~M.} \emph{et~al.}
\newblock \bibinfo{journal}{\bibinfo{title}{Strongly correlated quantum walks
  in optical lattices}}.
\newblock {\emph{\JournalTitle{Science}}} \textbf{\bibinfo{volume}{347}},
  \bibinfo{pages}{1229--1233}, \doiprefix\url{10.1126/science.1260364}
  (\bibinfo{year}{2015}).

\bibitem{Zakrzewski2017}
\bibinfo{author}{Wiater, D.}, \bibinfo{author}{Sowi\ifmmode~\acute{n}\else
  \'{n}\fi{}ski, T.} \& \bibinfo{author}{Zakrzewski, J.}
\newblock \bibinfo{journal}{\bibinfo{title}{Two bosonic quantum walkers in
  one-dimensional optical lattices}}.
\newblock {\emph{\JournalTitle{Phys. Rev. A}}} \textbf{\bibinfo{volume}{96}},
  \bibinfo{pages}{043629}, \doiprefix\url{10.1103/PhysRevA.96.043629}
  (\bibinfo{year}{2017}).

\bibitem{Chaohong_etc}
\bibinfo{author}{Qin, X.} \emph{et~al.}
\newblock \bibinfo{journal}{\bibinfo{title}{Statistics-dependent quantum
  co-walking of two particles in one-dimensional lattices with nearest-neighbor
  interactions}}.
\newblock {\emph{\JournalTitle{Phys. Rev. A}}} \textbf{\bibinfo{volume}{90}},
  \bibinfo{pages}{062301} (\bibinfo{year}{2014}).

\bibitem{ahlbrecht2012molecular}
\bibinfo{author}{Ahlbrecht, A.} \emph{et~al.}
\newblock \bibinfo{journal}{\bibinfo{title}{Molecular binding in interacting
  quantum walks}}.
\newblock {\emph{\JournalTitle{New Journal of Physics}}}
  \textbf{\bibinfo{volume}{14}}, \bibinfo{pages}{073050}
  (\bibinfo{year}{2012}).

\bibitem{mondalwalk}
\bibinfo{author}{Mondal, S.} \& \bibinfo{author}{Mishra, T.}
\newblock \bibinfo{journal}{\bibinfo{title}{Quantum walks of interacting
  mott-insulator defects with three-body interactions}}.
\newblock {\emph{\JournalTitle{Phys. Rev. A}}} \textbf{\bibinfo{volume}{101}},
  \doiprefix\url{10.1103/physreva.101.052341} (\bibinfo{year}{2020}).

\bibitem{Siloi2017}
\bibinfo{author}{Siloi, I.} \emph{et~al.}
\newblock \bibinfo{journal}{\bibinfo{title}{Noisy quantum walks of two
  indistinguishable interacting particles}}.
\newblock {\emph{\JournalTitle{Phys. Rev. A}}} \textbf{\bibinfo{volume}{95}},
  \bibinfo{pages}{022106}, \doiprefix\url{10.1103/PhysRevA.95.022106}
  (\bibinfo{year}{2017}).

\bibitem{Lahini_walk}
\bibinfo{author}{Lahini, Y.} \emph{et~al.}
\newblock \bibinfo{journal}{\bibinfo{title}{Quantum walk of two interacting
  bosons}}.
\newblock {\emph{\JournalTitle{Phys. Rev. A}}} \textbf{\bibinfo{volume}{86}},
  \bibinfo{pages}{011603}, \doiprefix\url{10.1103/PhysRevA.86.011603}
  (\bibinfo{year}{2012}).

\bibitem{poulios2014quantum}
\bibinfo{author}{Poulios, K.} \emph{et~al.}
\newblock \bibinfo{journal}{\bibinfo{title}{Quantum walks of correlated photon
  pairs in two-dimensional waveguide arrays}}.
\newblock {\emph{\JournalTitle{Phys. Rev. Lett.}}}
  \textbf{\bibinfo{volume}{112}}, \bibinfo{pages}{143604}
  (\bibinfo{year}{2014}).

\bibitem{zahringer2010realization}
\bibinfo{author}{Z{\"a}hringer, F.} \emph{et~al.}
\newblock \bibinfo{journal}{\bibinfo{title}{Realization of a quantum walk with
  one and two trapped ions}}.
\newblock {\emph{\JournalTitle{Phys. Rev. Lett.}}}
  \textbf{\bibinfo{volume}{104}}, \bibinfo{pages}{100503}
  (\bibinfo{year}{2010}).

\bibitem{Ye2019}
\bibinfo{author}{Ye, Y.} \emph{et~al.}
\newblock \bibinfo{journal}{\bibinfo{title}{Propagation and localization of
  collective excitations on a 24-qubit superconducting processor}}.
\newblock {\emph{\JournalTitle{Phys. Rev. Lett.}}}
  \textbf{\bibinfo{volume}{123}}, \bibinfo{pages}{050502},
  \doiprefix\url{10.1103/PhysRevLett.123.050502} (\bibinfo{year}{2019}).

\bibitem{Yan2019}
\bibinfo{author}{Yan, Z.} \emph{et~al.}
\newblock \bibinfo{journal}{\bibinfo{title}{Strongly correlated quantum walks
  with a 12-qubit superconducting processor}}.
\newblock {\emph{\JournalTitle{Science}}} \textbf{\bibinfo{volume}{364}},
  \bibinfo{pages}{753--756}, \doiprefix\url{10.1126/science.aaw1611}
  (\bibinfo{year}{2019}).

\bibitem{bloch_spin_charge_sep}
\bibinfo{author}{Vijayan, J.} \emph{et~al.}
\newblock \bibinfo{journal}{\bibinfo{title}{Time-resolved observation of
  spin-charge deconfinement in fermionic hubbard chains}}.
\newblock {\emph{\JournalTitle{Science}}} \textbf{\bibinfo{volume}{367}},
  \bibinfo{pages}{186}, \doiprefix\url{10.1126/science.aay2354}
  (\bibinfo{year}{2020}).

\bibitem{Kitagawa2012_topological_QW}
\bibinfo{author}{Kitagawa, T.} \emph{et~al.}
\newblock \bibinfo{journal}{\bibinfo{title}{Observation of topologically
  protected bound states in photonic quantum walks}}.
\newblock {\emph{\JournalTitle{Nature Communications}}}
  \textbf{\bibinfo{volume}{3}}, \bibinfo{pages}{882},
  \doiprefix\url{10.1038/ncomms1872} (\bibinfo{year}{2012}).

\bibitem{Ketterle2020}
\bibinfo{author}{Dimitrova, I.} \emph{et~al.}
\newblock \bibinfo{journal}{\bibinfo{title}{Enhanced superexchange in a tilted
  mott insulator}}.
\newblock {\emph{\JournalTitle{Phys. Rev. Lett.}}}
  \textbf{\bibinfo{volume}{124}}, \bibinfo{pages}{043204},
  \doiprefix\url{10.1103/PhysRevLett.124.043204} (\bibinfo{year}{2020}).

\bibitem{sowinskiprb2020}
\bibinfo{author}{Wrzosek, P.}, \bibinfo{author}{Wohlfeld, K.},
  \bibinfo{author}{Hofmann, D.}, \bibinfo{author}{Sowi{\'n}ski, T.} \&
  \bibinfo{author}{Sentef, M.~A.}
\newblock \bibinfo{journal}{\bibinfo{title}{Quantum walk versus classical wave:
  Distinguishing ground states of quantum magnets by spacetime dynamics}}.
\newblock {\emph{\JournalTitle{Phys. Rev. B}}} \textbf{\bibinfo{volume}{102}},
  \bibinfo{pages}{024440} (\bibinfo{year}{2020}).

\bibitem{Taglieber}
\bibinfo{author}{Taglieber, M.}, \bibinfo{author}{Voigt, A.-C.},
  \bibinfo{author}{Aoki, T.}, \bibinfo{author}{H\"ansch, T.~W.} \&
  \bibinfo{author}{Dieckmann, K.}
\newblock \bibinfo{journal}{\bibinfo{title}{Quantum degenerate two-species
  fermi-fermi mixture coexisting with a bose-einstein condensate}}.
\newblock {\emph{\JournalTitle{Phys. Rev. Lett.}}}
  \textbf{\bibinfo{volume}{100}}, \bibinfo{pages}{010401},
  \doiprefix\url{10.1103/PhysRevLett.100.010401} (\bibinfo{year}{2008}).

\bibitem{Fermi-Fermi}
\bibinfo{author}{Wille, E.} \emph{et~al.}
\newblock \bibinfo{journal}{\bibinfo{title}{Exploring an ultracold fermi-fermi
  mixture: Interspecies feshbach resonances and scattering properties of
  $^{6}\mathrm{Li}$ and $^{40}\mathrm{K}$}}.
\newblock {\emph{\JournalTitle{Phys. Rev. Lett.}}}
  \textbf{\bibinfo{volume}{100}}, \bibinfo{pages}{053201},
  \doiprefix\url{10.1103/PhysRevLett.100.053201} (\bibinfo{year}{2008}).

\bibitem{Ospelkaus2006}
\bibinfo{author}{Ospelkaus, S.} \emph{et~al.}
\newblock \bibinfo{journal}{\bibinfo{title}{Localization of bosonic atoms by
  fermionic impurities in a three-dimensional optical lattice}}.
\newblock {\emph{\JournalTitle{Phys. Rev. Lett.}}}
  \textbf{\bibinfo{volume}{96}}, \bibinfo{pages}{180403},
  \doiprefix\url{10.1103/PhysRevLett.96.180403} (\bibinfo{year}{2006}).

\bibitem{Tilman2006}
\bibinfo{author}{G\"unter, K.}, \bibinfo{author}{St\"oferle, T.},
  \bibinfo{author}{Moritz, H.}, \bibinfo{author}{K\"ohl, M.} \&
  \bibinfo{author}{Esslinger, T.}
\newblock \bibinfo{journal}{\bibinfo{title}{Bose-fermi mixtures in a
  three-dimensional optical lattice}}.
\newblock {\emph{\JournalTitle{Phys. Rev. Lett.}}}
  \textbf{\bibinfo{volume}{96}}, \bibinfo{pages}{180402},
  \doiprefix\url{10.1103/PhysRevLett.96.180402} (\bibinfo{year}{2006}).

\bibitem{Best}
\bibinfo{author}{Best, T.} \emph{et~al.}
\newblock \bibinfo{journal}{\bibinfo{title}{Role of interactions in
  $^{87}\mathrm{Rb}\mathrm{\text{\ensuremath{-}}}^{40}\mathbf{K}$ bose-fermi
  mixtures in a 3d optical lattice}}.
\newblock {\emph{\JournalTitle{Phys. Rev. Lett.}}}
  \textbf{\bibinfo{volume}{102}}, \bibinfo{pages}{030408},
  \doiprefix\url{10.1103/PhysRevLett.102.030408} (\bibinfo{year}{2009}).

\bibitem{Catani}
\bibinfo{author}{Catani, J.}, \bibinfo{author}{De~Sarlo, L.},
  \bibinfo{author}{Barontini, G.}, \bibinfo{author}{Minardi, F.} \&
  \bibinfo{author}{Inguscio, M.}
\newblock \bibinfo{journal}{\bibinfo{title}{Degenerate bose-bose mixture in a
  three-dimensional optical lattice}}.
\newblock {\emph{\JournalTitle{Phys. Rev. A}}} \textbf{\bibinfo{volume}{77}},
  \bibinfo{pages}{011603}, \doiprefix\url{10.1103/PhysRevA.77.011603}
  (\bibinfo{year}{2008}).

\bibitem{Gadway}
\bibinfo{author}{Gadway, B.}, \bibinfo{author}{Pertot, D.},
  \bibinfo{author}{Reimann, R.} \& \bibinfo{author}{Schneble, D.}
\newblock \bibinfo{journal}{\bibinfo{title}{Superfluidity of interacting
  bosonic mixtures in optical lattices}}.
\newblock {\emph{\JournalTitle{Phys. Rev. Lett.}}}
  \textbf{\bibinfo{volume}{105}}, \bibinfo{pages}{045303},
  \doiprefix\url{10.1103/PhysRevLett.105.045303} (\bibinfo{year}{2010}).

\bibitem{Altman2003}
\bibinfo{author}{Altman, E.}, \bibinfo{author}{Hofstetter, W.},
  \bibinfo{author}{Demler, E.} \& \bibinfo{author}{Lukin, M.~D.}
\newblock \bibinfo{journal}{\bibinfo{title}{Phase diagram of two-component
  bosons on an optical lattice}}.
\newblock {\emph{\JournalTitle{New Journal of Physics}}}
  \textbf{\bibinfo{volume}{5}}, \bibinfo{pages}{113--113},
  \doiprefix\url{10.1088/1367-2630/5/1/113} (\bibinfo{year}{2003}).

\bibitem{Isacsson2005}
\bibinfo{author}{Isacsson, A.}, \bibinfo{author}{Cha, M.-C.},
  \bibinfo{author}{Sengupta, K.} \& \bibinfo{author}{Girvin, S.~M.}
\newblock \bibinfo{journal}{\bibinfo{title}{Superfluid-insulator transitions of
  two-species bosons in an optical lattice}}.
\newblock {\emph{\JournalTitle{Phys. Rev. B}}} \textbf{\bibinfo{volume}{72}},
  \bibinfo{pages}{184507}, \doiprefix\url{10.1103/PhysRevB.72.184507}
  (\bibinfo{year}{2005}).

\bibitem{Duan2003}
\bibinfo{author}{Duan, L.-M.}, \bibinfo{author}{Demler, E.} \&
  \bibinfo{author}{Lukin, M.~D.}
\newblock \bibinfo{journal}{\bibinfo{title}{Controlling spin exchange
  interactions of ultracold atoms in optical lattices}}.
\newblock {\emph{\JournalTitle{Phys. Rev. Lett.}}}
  \textbf{\bibinfo{volume}{91}}, \bibinfo{pages}{090402},
  \doiprefix\url{10.1103/PhysRevLett.91.090402} (\bibinfo{year}{2003}).

\bibitem{Orth2008}
\bibinfo{author}{Orth, P.~P.}, \bibinfo{author}{Stanic, I.} \&
  \bibinfo{author}{Le~Hur, K.}
\newblock \bibinfo{journal}{\bibinfo{title}{Dissipative quantum ising model in
  a cold-atom spin-boson mixture}}.
\newblock {\emph{\JournalTitle{Phys. Rev. A}}} \textbf{\bibinfo{volume}{77}},
  \bibinfo{pages}{051601}, \doiprefix\url{10.1103/PhysRevA.77.051601}
  (\bibinfo{year}{2008}).

\bibitem{Wang2016}
\bibinfo{author}{Wang, W.}, \bibinfo{author}{Penna, V.} \&
  \bibinfo{author}{Capogrosso-Sansone, B.}
\newblock \bibinfo{journal}{\bibinfo{title}{Inter-species entanglement of
  bose{\textendash}bose mixtures trapped in optical lattices}}.
\newblock {\emph{\JournalTitle{New Journal of Physics}}}
  \textbf{\bibinfo{volume}{18}}, \bibinfo{pages}{063002},
  \doiprefix\url{10.1088/1367-2630/18/6/063002} (\bibinfo{year}{2016}).

\bibitem{Mathey2007}
\bibinfo{author}{Mathey, L.}
\newblock \bibinfo{journal}{\bibinfo{title}{Commensurate mixtures of ultracold
  atoms in one dimension}}.
\newblock {\emph{\JournalTitle{Phys. Rev. B}}} \textbf{\bibinfo{volume}{75}},
  \bibinfo{pages}{144510}, \doiprefix\url{10.1103/PhysRevB.75.144510}
  (\bibinfo{year}{2007}).

\bibitem{mishraps}
\bibinfo{author}{Mishra, T.}, \bibinfo{author}{Pai, R.~V.} \&
  \bibinfo{author}{Das, B.~P.}
\newblock \bibinfo{journal}{\bibinfo{title}{Phase separation in a two-species
  bose mixture}}.
\newblock {\emph{\JournalTitle{Phys. Rev. A}}} \textbf{\bibinfo{volume}{76}},
  \bibinfo{pages}{013604}, \doiprefix\url{10.1103/PhysRevA.76.013604}
  (\bibinfo{year}{2007}).

\bibitem{Singh2017}
\bibinfo{author}{Singh, M.}, \bibinfo{author}{Mondal, S.},
  \bibinfo{author}{Sahoo, B.~K.} \& \bibinfo{author}{Mishra, T.}
\newblock \bibinfo{journal}{\bibinfo{title}{Quantum phases of constrained
  dipolar bosons in coupled one-dimensional optical lattices}}.
\newblock {\emph{\JournalTitle{Phys. Rev. A}}} \textbf{\bibinfo{volume}{96}},
  \bibinfo{pages}{053604}, \doiprefix\url{10.1103/PhysRevA.96.053604}
  (\bibinfo{year}{2017}).

\bibitem{SSHHexptLe2020}
\bibinfo{author}{Le, N.~H.}, \bibinfo{author}{Fisher, A.~J.},
  \bibinfo{author}{Curson, N.~J.} \& \bibinfo{author}{Ginossar, E.}
\newblock \bibinfo{journal}{\bibinfo{title}{Topological phases of a dimerized
  fermi--hubbard model for semiconductor nano-lattices}}.
\newblock {\emph{\JournalTitle{npj Quantum Information}}}
  \textbf{\bibinfo{volume}{6}}, \bibinfo{pages}{24},
  \doiprefix\url{10.1038/s41534-020-0253-9} (\bibinfo{year}{2020}).

\bibitem{SantosSSH2018}
\bibinfo{author}{Barbiero, L.}, \bibinfo{author}{Santos, L.} \&
  \bibinfo{author}{Goldman, N.}
\newblock \bibinfo{journal}{\bibinfo{title}{Quenched dynamics and spin-charge
  separation in an interacting topological lattice}}.
\newblock {\emph{\JournalTitle{Phys. Rev. B}}} \textbf{\bibinfo{volume}{97}},
  \bibinfo{pages}{201115}, \doiprefix\url{10.1103/PhysRevB.97.201115}
  (\bibinfo{year}{2018}).

\bibitem{Mondal2020}
\bibinfo{author}{Mondal, S.}, \bibinfo{author}{Greschner, S.},
  \bibinfo{author}{Santos, L.} \& \bibinfo{author}{Mishra, T.}
\newblock \bibinfo{journal}{\bibinfo{title}{Topological inheritance in
  two-component hubbard models with single-component su-schrieffer-heeger
  dimerization}}.
\newblock {\emph{\JournalTitle{Phys. Rev. A}}} \textbf{\bibinfo{volume}{104}},
  \bibinfo{pages}{013315}, \doiprefix\url{10.1103/PhysRevA.104.013315}
  (\bibinfo{year}{2021}).

\bibitem{Ye2016}
\bibinfo{author}{Ye, B.-T.}, \bibinfo{author}{Mu, L.-Z.} \&
  \bibinfo{author}{Fan, H.}
\newblock \bibinfo{journal}{\bibinfo{title}{Entanglement spectrum of
  su-schrieffer-heeger-hubbard model}}.
\newblock {\emph{\JournalTitle{Phys. Rev. B}}} \textbf{\bibinfo{volume}{94}},
  \bibinfo{pages}{165167}, \doiprefix\url{10.1103/PhysRevB.94.165167}
  (\bibinfo{year}{2016}).

\bibitem{Mandel_et_al2003}
\bibinfo{author}{Mandel, O.} \emph{et~al.}
\newblock \bibinfo{journal}{\bibinfo{title}{Coherent transport of neutral atoms
  in spin-dependent optical lattice potentials}}.
\newblock {\emph{\JournalTitle{Phys. Rev. Lett.}}}
  \textbf{\bibinfo{volume}{91}}, \bibinfo{pages}{010407},
  \doiprefix\url{10.1103/PhysRevLett.91.010407} (\bibinfo{year}{2003}).

\bibitem{Soltan-Panahi2011}
\bibinfo{author}{Soltan-Panahi, P.} \emph{et~al.}
\newblock \bibinfo{journal}{\bibinfo{title}{Multi-component quantum gases in
  spin-dependent hexagonal lattices}}.
\newblock {\emph{\JournalTitle{Nature Physics}}} \textbf{\bibinfo{volume}{7}},
  \bibinfo{pages}{434--440}, \doiprefix\url{10.1038/nphys1916}
  (\bibinfo{year}{2011}).

\bibitem{Jian-wei-2017}
\bibinfo{author}{Yang, B.} \emph{et~al.}
\newblock \bibinfo{journal}{\bibinfo{title}{Spin-dependent optical
  superlattice}}.
\newblock {\emph{\JournalTitle{Phys. Rev. A}}} \textbf{\bibinfo{volume}{96}},
  \bibinfo{pages}{011602}, \doiprefix\url{10.1103/PhysRevA.96.011602}
  (\bibinfo{year}{2017}).

\bibitem{Esslinger2015PRL}
\bibinfo{author}{Jotzu, G.} \emph{et~al.}
\newblock \bibinfo{journal}{\bibinfo{title}{Creating state-dependent lattices
  for ultracold fermions by magnetic gradient modulation}}.
\newblock {\emph{\JournalTitle{Phys. Rev. Lett.}}}
  \textbf{\bibinfo{volume}{115}}, \bibinfo{pages}{073002},
  \doiprefix\url{10.1103/PhysRevLett.115.073002} (\bibinfo{year}{2015}).

\bibitem{blochimbalance}
\bibinfo{author}{Oppong, N.~D.} \emph{et~al.}
\newblock \bibinfo{journal}{\bibinfo{title}{Probing transport and slow
  relaxation in the mass-imbalanced fermi-hubbard model}}.
\newblock {\emph{\JournalTitle{arXiv:2011.12411}}}  (\bibinfo{year}{2020}).

\bibitem{Roos2017}
\bibinfo{author}{Roos, C.~F.}, \bibinfo{author}{Alberti, A.},
  \bibinfo{author}{Meschede, D.}, \bibinfo{author}{Hauke, P.} \&
  \bibinfo{author}{H\"affner, H.}
\newblock \bibinfo{journal}{\bibinfo{title}{Revealing quantum statistics with a
  pair of distant atoms}}.
\newblock {\emph{\JournalTitle{Phys. Rev. Lett.}}}
  \textbf{\bibinfo{volume}{119}}, \bibinfo{pages}{160401},
  \doiprefix\url{10.1103/PhysRevLett.119.160401} (\bibinfo{year}{2017}).

\bibitem{LindaSansoni2012}
\bibinfo{author}{Sansoni, L.} \emph{et~al.}
\newblock \bibinfo{journal}{\bibinfo{title}{Two-particle bosonic-fermionic
  quantum walk via integrated photonics}}.
\newblock {\emph{\JournalTitle{Phys. Rev. Lett.}}}
  \textbf{\bibinfo{volume}{108}}, \bibinfo{pages}{010502},
  \doiprefix\url{10.1103/PhysRevLett.108.010502} (\bibinfo{year}{2012}).

\bibitem{Aidelsburger2018}
\bibinfo{author}{Scherg, S.} \emph{et~al.}
\newblock \bibinfo{journal}{\bibinfo{title}{Nonequilibrium mass transport in
  the 1d fermi-hubbard model}}.
\newblock {\emph{\JournalTitle{Phys. Rev. Lett.}}}
  \textbf{\bibinfo{volume}{121}}, \bibinfo{pages}{130402},
  \doiprefix\url{10.1103/PhysRevLett.121.130402} (\bibinfo{year}{2018}).

\bibitem{Sowinski_sarkar}
\bibinfo{author}{Sarkar, S.} \& \bibinfo{author}{Sowi\ifmmode~\acute{n}\else
  \'{n}\fi{}ski, T.}
\newblock \bibinfo{journal}{\bibinfo{title}{Correlations in few two-component
  quantum walkers on a tilted lattice}}.
\newblock {\emph{\JournalTitle{Phys. Rev. A}}} \textbf{\bibinfo{volume}{102}},
  \bibinfo{pages}{043326}, \doiprefix\url{10.1103/PhysRevA.102.043326}
  (\bibinfo{year}{2020}).

\bibitem{Winkler2006}
\bibinfo{author}{Winkler, K.} \emph{et~al.}
\newblock \bibinfo{journal}{\bibinfo{title}{Repulsively bound atom pairs in an
  optical lattice}}.
\newblock {\emph{\JournalTitle{Nature}}} \textbf{\bibinfo{volume}{441}},
  \bibinfo{pages}{853--856}, \doiprefix\url{10.1038/nature04918}
  (\bibinfo{year}{2006}).

\bibitem{Farhi1998}
\bibinfo{author}{Farhi, E.} \& \bibinfo{author}{Gutmann, S.}
\newblock \bibinfo{journal}{\bibinfo{title}{Quantum computation and decision
  trees}}.
\newblock {\emph{\JournalTitle{Phys. Rev. A}}} \textbf{\bibinfo{volume}{58}},
  \bibinfo{pages}{915--928}, \doiprefix\url{10.1103/PhysRevA.58.915}
  (\bibinfo{year}{1998}).

\bibitem{Kempe_2003}
\bibinfo{author}{Kempe, J.}
\newblock \bibinfo{journal}{\bibinfo{title}{Quantum random walks: An
  introductory overview}}.
\newblock {\emph{\JournalTitle{Contemporary Physics}}}
  \textbf{\bibinfo{volume}{44}}, \bibinfo{pages}{307–327},
  \doiprefix\url{10.1080/00107151031000110776} (\bibinfo{year}{2003}).

\bibitem{Childs_2002}
\bibinfo{journal}{\bibinfo{author}{Childs, A.~M.}, \bibinfo{author}{Farhi, E.}
  \& \bibinfo{author}{Gutmann, S.}}
\newblock {\emph{\JournalTitle{Quantum Information Processing}}}
  \textbf{\bibinfo{volume}{1}}, \bibinfo{pages}{35–43},
  \doiprefix\url{10.1023/a:1019609420309} (\bibinfo{year}{2002}).

\bibitem{scirep_3B}
\bibinfo{author}{Ren, J.}, \bibinfo{author}{Wu, Y.-Z.} \& \bibinfo{author}{Xu,
  X.-F.}
\newblock \bibinfo{journal}{\bibinfo{title}{Expansion dynamics in a
  one-dimensional hard-core boson model with three-body interactions}}.
\newblock {\emph{\JournalTitle{Scientific Reports}}}
  \textbf{\bibinfo{volume}{5}}, \bibinfo{pages}{14743},
  \doiprefix\url{10.1038/srep14743} (\bibinfo{year}{2015}).

\bibitem{Sowi_ski_2019}
\bibinfo{author}{Sowi{\'{n}}ski, T.} \& \bibinfo{author}{Garc{\'{\i}}a-March,
  M.~{\'{A}}.}
\newblock \bibinfo{journal}{\bibinfo{title}{One-dimensional mixtures of several
  ultracold atoms: a review}}.
\newblock {\emph{\JournalTitle{Reports on Progress in Physics}}}
  \textbf{\bibinfo{volume}{82}}, \bibinfo{pages}{104401},
  \doiprefix\url{10.1088/1361-6633/ab3a80} (\bibinfo{year}{2019}).

\bibitem{PhysRevB.78.184513}
\bibinfo{author}{Kalas, R.~M.}, \bibinfo{author}{Balatsky, A.~V.} \&
  \bibinfo{author}{Mozyrsky, D.}
\newblock \bibinfo{journal}{\bibinfo{title}{Odd-frequency pairing in a binary
  mixture of bosonic and fermionic cold atoms}}.
\newblock {\emph{\JournalTitle{Phys. Rev. B}}} \textbf{\bibinfo{volume}{78}},
  \bibinfo{pages}{184513}, \doiprefix\url{10.1103/PhysRevB.78.184513}
  (\bibinfo{year}{2008}).

\bibitem{Duan_altman2003}
\bibinfo{author}{Duan, L.-M.}, \bibinfo{author}{Demler, E.} \&
  \bibinfo{author}{Lukin, M.~D.}
\newblock \bibinfo{journal}{\bibinfo{title}{Controlling spin exchange
  interactions of ultracold atoms in optical lattices}}.
\newblock {\emph{\JournalTitle{Phys. Rev. Lett.}}}
  \textbf{\bibinfo{volume}{91}}, \bibinfo{pages}{090402},
  \doiprefix\url{10.1103/PhysRevLett.91.090402} (\bibinfo{year}{2003}).

\end{thebibliography}

\section*{Acknowledgments}
The computational simulations were carried out using the Param-Ishan HPC facility at Indian Institute of Technology - Guwahati, India. T.M. acknowledges DST-SERB, India for the early career grant through Project No.ECR/2017/001069.

\section*{Author contributions}
T.M. and B.P.D. planned the research; M.K.G., S.M. performed the calculation, M.K.G, S.M. B.P.D. and T.M. interpreted the results; S.M, B.P.D and T.M. wrote the paper. T.M. supervised the research.

\section*{Competing interests}
The authors declare no competing interests.

\section*{Additional information}
Correspondence and requests for materials should be addressed to T.M. or B.P.D.
\end{document}